\documentclass[8.5pt,twoside,twocolumn]{article}
\usepackage[super,sort&compress,comma]{natbib}
\usepackage[version=3]{mhchem} % Formula subscripts using \ce{}
\usepackage{times,mathptmx}
\usepackage{sectsty}
\usepackage{balance}

\usepackage{graphicx} %eps figures can be used instead
\usepackage{lastpage}
\usepackage[format=plain,justification=raggedright,singlelinecheck=false,font=small,labelfont=bf,labelsep=space]{caption}
\usepackage{fancyhdr}
\usepackage{color}
\usepackage{extarrows}
\newcommand{\si}{ESI\dag}

\newcommand{\ib}[1]{{\color{black}#1}}
\newcommand{\last}[1]{{\color{blue}#1}}

\begin{document}

\thispagestyle{plain}
\fancypagestyle{plain}{
%13.09.2014 \fancyhead[L]{\includegraphics[height=8pt]{headers/LH}}
% \fancyhead[C]{\hspace{-1cm}\includegraphics[height=20pt]{headers/CH}}
%13.09.2014 \fancyhead[R]{\includegraphics[height=10pt]{headers/RH}\vspace{-0.2cm}}
\renewcommand{\headrulewidth}{1pt}}
\renewcommand{\thefootnote}{\fnsymbol{footnote}}
\renewcommand\footnoterule{\vspace*{1pt}%
\hrule width 3.4in height 0.4pt \vspace*{5pt}}
\setcounter{secnumdepth}{5}

\makeatletter
\def\subsubsection{\@startsection{subsubsection}{3}{10pt}{-1.25ex plus -1ex minus -.1ex}{0ex plus 0ex}{\normalsize\bf}}
\def\paragraph{\@startsection{paragraph}{4}{10pt}{-1.25ex plus -1ex minus -.1ex}{0ex plus 0ex}{\normalsize\textit}}
\renewcommand\@biblabel[1]{#1}
\renewcommand\@makefntext[1]%
{\noindent\makebox[0pt][r]{\@thefnmark\,}#1}
\makeatother
\renewcommand{\figurename}{\small{Fig.}~}
\sectionfont{\large}
\subsectionfont{\normalsize}

\fancyfoot{}
%13.09.2014 \fancyfoot[LO,RE]{\vspace{-7pt}\includegraphics[height=9pt]{headers/LF}}
%13.09.2014 \fancyfoot[CO]{\vspace{-7.2pt}\hspace{12.2cm}\includegraphics{headers/RF}}
%13.09.2014 \fancyfoot[CE]{\vspace{-7.5pt}\hspace{-13.5cm}\includegraphics{headers/RF}}
\fancyfoot[RO]{\footnotesize{\sffamily{1--\pageref{LastPage} ~\textbar  \hspace{2pt}\thepage}}}
\fancyfoot[LE]{\footnotesize{\sffamily{\thepage~\textbar\hspace{3.45cm} 1--\pageref{LastPage}}}}
\fancyhead{}
\renewcommand{\headrulewidth}{1pt}
\renewcommand{\footrulewidth}{1pt}
\setlength{\arrayrulewidth}{1pt}
\setlength{\columnsep}{6.5mm}
\setlength\bibsep{1pt}

\newcommand{\alt}{\raisebox{-0.3ex}{$\stackrel{<}{\sim}$}}
\newcommand{\agt}{\raisebox{-0.3ex}{$\stackrel{>}{\sim}$}}
% \graphicspath{{ }{.}{../}{../../}{./kai/}}
\DeclareGraphicsExtensions{.eps,.ps}

\twocolumn[
  \begin{@twocolumnfalse}
\noindent\LARGE{\textbf{Important issues facing model-based approaches to tunneling transport in molecular junctions $^\dag$}}
\vspace{0.6cm}

\noindent\large{\textbf{Ioan B\^aldea $^{\ast}$
\textit{$^{a\ddag}$}
}}\vspace{0.5cm}

\noindent \textbf{\small{Published: Phys.~Chem.~Chem.~Phys.~2015, {\bf 17}, 20217-20230; DOI: 10.1039/C5CP02595H}}
\vspace{0.6cm}

\noindent
\normalsize{Abstract:\\
Extensive studies on thin films indicated a generic cubic current-voltage $I-V$ dependence as a salient feature of charge transport by tunneling. A quick glance at $I-V$ data for molecular junctions suggests a qualitatively similar behavior. This would render model-based studies almost irrelevant, since, whatever the model, its parameters can always be adjusted to fit symmetric (asymmetric) $I-V$ curves characterized by two (three) expansion coefficients. Here, we systematically examine popular models based on tunneling barrier or tight-binding pictures and demonstrate that, for a quantitative description at biases of interest ($V$ slightly higher than the transition voltage $V_t$), cubic expansions do not suffice. A detailed collection of analytical formulae as well as their conditions of applicability 
are presented to facilitate experimentalists colleagues 
to process and interpret their experimental data by obtained by measuring currents in molecular junctions. 
We discuss in detail the limits of applicability of the various models and emphasize that uncritically adjusting model parameters to experiment may be unjustified because the values deduced in this way may fall in ranges rendering a specific model invalid or incompatible to ab initio estimates. We exemplify with the benchmark case of oligophenylene-based junctions, for which results of ab initio quantum chemical calculations are also reported. As a specific issue, we address the impact of the spatial potential profile and show that it is not notable up to biases $V \agt V_t$, unlike at higher biases, where it may be responsible for negative differential resistance effects.

$ $ \\

}
\vspace{0.5cm}
 \end{@twocolumnfalse}
  ]

%Footnotes
%Please use \dag to cite the ESI in the main text of the article.
%If you article does not have ESI please remove the the \dag symbol from the title and the above footnotetext.

\footnotetext{\textit{$^{a}$~Theoretische Chemie, Universit\"at Heidelberg, Im Neuenheimer Feld 229, D-69120 Heidelberg, Germany.}}
\footnotetext{\dag~Electronic supplementary information (ESI) available. See DOI: 10.1039/C5CP02595H}
\footnotetext{\ddag~E-mail: ioan.baldea@pci.uni-heidelberg.de.
Also at National Institute for Lasers, Plasmas, and Radiation Physics, Institute of Space Sciences,
RO 077125, Bucharest-M\u{a}gurele, Romania}
%
% 
%%%%%%%%%%%%%%%%%%%%%%%%%%%%%%%%%%%%%%%%%%%%%%%%%%%%%%%%%%%%%%%%%%%%%%%%%%%%%%%%%%%%%%%%%%%%%%%%% 
%  -------------------------
%%%%%%%%%%%%%%%%%%%%%%%%%%%%%%%%%%%%%%%%%%%%%%%%%%%%%%%%%%%%%%%%%%%%%%%%%%%%%%%%%%%%%%%%%%%%%%%%% 
\section{Introduction and background}
\label{sec:intro}
The roughly parabolic shape of the conductance $G(V) \equiv \partial I/\partial V$, 
or the related cubic dependence of the current 
($I$) on bias ($V$), was
considered a prominent characteristic of transport via tunneling. This conclusion  emerged from 
extensive studies on a variety of macroscopic
thin film junctions of oxides, insulators, superconductors up to relatively large biases.
\cite{Simmons:63,Simmons:63d,Rowell:69,Duke:69b,Duke:69,Brinkman:70} 
A quick glance at $I-V$ measurements in a variety of molecular junctions 
may convey the impression that this cubic dependence is satisfactory also for such systems.\cite{CuevasScheer:10}
As illustration, we have chosen in \figurename\ref{fig:generic-parabola}a a measured $I-V$ curve,\cite{baldea:2015,Baldea:2015d}
for which fitting with a cubic polynomial looks particularly accurate.
The view based of such third-order Taylor expansions (cf.~eqn~(\ref{eq-i3})) was  
able to qualitatively describe a series of interesting aspects related to 
charge transport in molecular junctions \cite{Vilan:13}
%%%%%%%%%%%%%%%%%%%%%%%%%%%%%%%%%%%%%%%%%%%%%%%%%%%%%%%%%%%%%%%%%%%%%%%%%%%%%%%%%%%%%%%%%%%%%%%%% 
\begin{eqnarray}
I(V) & = & I_3 (V) + \mathcal{O}\left(V^5\right);  I_3 (V) \equiv G \left( V + c_{2} V^3\right) 
\nonumber \\
G(V) & \equiv & \frac{\partial}{\partial V} I(V) = G\left(1 + 3 c_{2} V^2\right) + \mathcal{O}\left(V^4\right) 
\label{eq-i3}
\end{eqnarray}
%%%%%%%%%%%%%%%%%%%%%%%%%%%%%%%%%%%%%%%%%%%%%%%%%%%%%%%%%%%%%%%%%%%%%%%%%%%%%%%%%%%%%%%%%%%%%%%%% 
Above, $G \equiv \lim_{V\to 0} I/V$ is the low bias conductance.\cite{omit-even-powers}

However, if the approximation of eqn~(\ref{eq-i3}) were \emph{quantitatively} adequate, it would be 
deceptive for a model-based description of transport in molecular junctions. 
Resorting to simple phenomenological models enabling expedient data processing and interpretation of charge 
transport through nanoscale devices represents common practice 
in molecular electronics. 
% \ib
{The validity of a certain model/mechanism for transport in a given system/device 
is often assessed
by considering its ability to fit the measured current-voltage $I-V$ curves.}
Adopting this ``pragmatic'' standpoint, the problem encountered is that an approximate generic parabolic shape of the conductance 
could be inferred at not too high biases 
regardless the tunneling model.\cite{CuevasScheer:10,Vilan:13} 
Whether the electron wave function tunnels across 
a structureless average medium modeled as a tunneling barrier 
or through tails of densities of states of one (frontier) or a few off-resonant molecular orbital levels, 
one could ``appropriately'' adjust a few 
parameters, and the third-order expansion of the current formula $I=I(V)$,  often available in closed analytical forms 
from literature,\cite{Simmons:63,Rowell:69,Duke:69b,Brinkman:70,Mujica:01,Baldea:2012a}
could relatively easy be ``made'' to fit the usually featureless $I-V$ curves measured.
Then the quality of the fitting alone cannot be invoked in favor of one tunneling mechanism out of other 
possible mechanisms underlying the various phenomenological models. 

% \ib
{
The inspection of \figurename\ref{fig:generic-parabola}a may indeed convey the impression of an 
overall very good agreement between experiment and eqn~(\ref{eq-i3}).
However, a more careful analysis reveals that 
the range of the highest biases that can be sampled experimentally is 
quantitatively not so satisfactorily described.
Typically, these $V$-values are only slightly larger than 
the transition voltage $V_t$.\cite{Beebe:06} 
% an important characteristic property of molecular junctions (cf.~Section~\ref{sec:Vt}). 
The difference in $V_t$-values 
extracted from the experimental and fitting curves of 
\figurename\ref{fig:generic-parabola}b exceeds typical $V_t$-experimental errors.
%%%%%%%%%%%%%%%%%%%%%%%%%%%%%%%%%%%%%%%%%%%%%%%%%%%%%%%%%%%%%%%%%%%%%%%%%%%%%%%%%%%%%%%%%%%%%%%%%%%%%%% 
\begin{figure*}[htb]
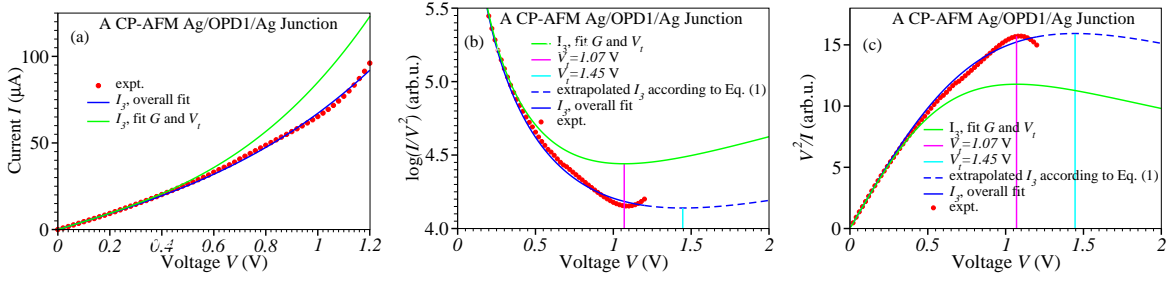

\centerline{\includegraphics[width=0.28\textwidth,angle=0]{fig_iv_AgAg_1B_11_g_4.53927071428572e-05_i0_8.04742142857143e-07_c2_0.478291197010046.eps}
\hspace*{0ex}
\includegraphics[width=0.28\textwidth,angle=0]{fig_fn_AgAg_1B_11_g_4.53927071428572e-05_i0_8.04742142857143e-07_c2_0.478291197010046.eps}
\hspace*{0ex}
\includegraphics[width=0.28\textwidth,angle=0]{fig_ib_AgAg_1B_11_g_4.53927071428572e-05_i0_8.04742142857143e-07_c2_0.478291197010046.eps}
}
% $ $\\[0ex]
\caption{(a) Fitting the red $I-V$ curve measured for a CP-AFM Ag/OPD1/Ag molecular junction 
% (courtesy of Zuoti Xie) 
\cite{baldea:2015}
with a cubic polynomial, eqn~(\ref{eq-i3}), depicted by the blue line may appear to be quite satisfactory. (b) Nevertheless, for the same curves, the difference in the positions of the minimum of the Fowler-Nordheim (FN) quantity $\log\left(I/V^2\right)$ defining the transition voltage $V_t$ is substantially larger than experimental uncertainties ($\delta V_t < 0.1$\,V \cite{baldea:2015}). In the case illustrated here, 
the transition voltage obtained via eqn~(\ref{eq-i3}) is even beyond the bias range accessed experimentally. (c) As alternative to the FN plot, one can inspect a ``peak voltage spectrum'', namely 
the quantity $V^2/I$ plotted vs.~$V$,\cite{pvs} which exhibits a maximum located at $V=V_t$. 
(We show only the range $V>0$ because the measured $I-V$ curve is 
symmetric to a very good approximation.)}     
\label{fig:generic-parabola}
\end{figure*}
%%%%%%%%%%%%%%%%%%%%%%%%%%%%%%%%%%%%%%%%%%%%%%%%%%%%%%%%%%%%%%%%%%%%%%%%%%%%%%%%%%%%%%%%%%%%%%%%%%%%%%% 

The range $V\sim V_t$ is of special interest. As discussed in Section \ref{sec:Vt}, 
$V_t$ represents an intrinsic property of a junction out of equilibrium;
it characterizes the (high) bias range exhibiting significant nonlinearity.
Therefore, even more than an overall fitting of $I-V$ transport data,
it is important for theory to correctly understand/reproduce $V_t$. 
And, as illustrated in \figurename\ref{fig:generic-parabola}b and c, 
to this aim, the theoretical description should go beyond the cubic expansion framework provided by eqn~(\ref{eq-i3}).
Emphasizing this aspect in the analysis of the various transport models considered below,
which are among the most popular in the molecular electronics community,
\cite{Simmons:63,Simmons:63d,Gundlach:69b,Duke:69,Rowell:69,Brinkman:70,Nitzan:01,Mujica:01,Araidai:10} 
represents one main aim of the present study.
}

Emerging from idealized description of reality, 
the models utilized in transport studies, like those 
used elsewhere, 
% \ib
{inherently} 
have a limited validity. 
The various model parameter values are subject to specific restrictions,
which represent \emph{intrinsic} limitations for the model in question; 
merely succeeding to provide good quality data fitting is meaningless if the
the parameter values deduced in this way fall outside the range of model's validity.
In some cases, these conditions of validity simply reflect 
restrictions on parameter values justifying certain mathematical approximations
made to express a certain result in closed analytical form.  
Checking whether the various parameters deduced from data fitting do indeed satisfy  
the corresponding (mathematical) restrictions or are consistent with ab initio estimated values
is absolutely necessary. 
Nevertheless, not rarely current data analysis misses this important consistency step, which may be related
to an insufficient discussion in the literature
of the physical background of the various phenomenological 
models and the limited validity of the pertaining analytical formulas. 
% \ib
{Exposing the limits of applicability of such models frequently used in molecular electronics represents another aim of the present paper.
In doing that, this paper is intended as an effort to make the community more aware of these limitations. 
It should by no means be understood as challenging the utility of model-based studies
in gaining valuable conceptual physical insight into charge transport.}

Throughout, we will consider the case of zero temperature, as appropriate for off-resonant tunneling and symmetric 
coupling to electrodes (equal width parameters $\Gamma_L = \Gamma_R = \Gamma$ due to ``left'' and ``right'' electrodes). In agreement with the fact that conductances of typical molecular junctions are much smaller 
than the quantum conductance $G \ll G_0 = 2 e^2/h = 77.48\,\mu$S, we will assume throughout $\Gamma$'s 
much smaller than relevant (molecular orbital) energy offsets relative electrodes' Fermi energy 
$\varepsilon_B \equiv E_B - E_F$ ($E_F \equiv 0$).
We will further assume $\varepsilon_B > 0$, because, for tunneling barriers, 
the corresponding formulas are simpler to write for electrons (n-type conduction); 
in all formulas presented below,
$\varepsilon_B$ should be understood as $\vert\varepsilon_B\vert$ 
in cases of holes (p-type conduction) 
tunneling across negative barriers, as easily obtained by performing the charge conjugation
transformation ($\varepsilon_B = E_B - E_F \to E_F - E_B = - \varepsilon_B$, $e \to -e$, $m\to -m$, 
see, e.g., ref.~\citenum{Baldea:2012c}). Likewise, since the tight-binding Hamiltonians of 
% eqn~(\ref{eq-ham-chains}) and (\ref{eq-ham-rings}) 
eqn~\ib{(S1)} and \ib{(S3)} 
are invariant under particle-hole 
conjugation (e.g. ref.~\citenum{Baldea:2004a} and citations therein), 
the results for $+\varepsilon_B$ and $-\varepsilon_B$ coincide.

While discussing in general the aspects delineated above, 
we will examine the benchmark case of conducting probe atomic
force microscope (CP-AFM) molecular junctions based on oligophenylene 
dithiols (OPDs) 
% of a recent study
\cite{baldea:2015} as a specific example.
%
% \ib{To avoid possible misunderstandings, a comment in the end of this section is in order, however.
% By exposing the limitations of the various models considered below, the present paper is indended as an effort 
% to make the community more aware of them and by no means to challenge the utility of model-based studies
% in gaining valuable conceptual physical insight.}
%%%%%%%%%%%%%%%%%%%%%%%%%%%%%%%%%%%%%%%%%%%%%%%%%%%%%%%%%%%%%%%%%%%%%%%%%%%%%%%%%%%%%%%%%%%%%%%%%
\section{Physical properties envisaged}
\label{sec:Vt} 
%%%%%%%%%%%%%%%%%%%%%%%%%%%%%%%%%%%%%%%%%%%%%%%%%%%%%%%%%%%%%%%%%%%%%%%%%%%%%%%%%%%%%%%%%%%%%%%%%
In this paper, we will mainly consider two physical properties (zero-bias conductance
$G$ and transition voltage $V_t$  \cite{Beebe:06}) and focus our attention
on how these properties vary with the molecular length $d$ or size $N$ across 
homologous series of molecules described within schematic models. 
Typical for nonresonant tunneling transport is 
a dependence 
%%%%%%%%%%%%%%%%%%%%%%%%%%%%%%%%%%%%%%%%%%%%%%%%%%%%%%%%%%%%
\begin{equation}
\label{eq-G-n}
G \propto \exp(-\beta N) = \exp\left( - \tilde{\beta} d\right)
\end{equation}
%%%%%%%%%%%%%%%%%%%%%%%%%%%%%%%%%%%%%%%%%%%%%%%%%%%%%%%%%%%%
where $\beta$ ($\tilde{\beta}$) characterizes the exponential decay 
of the zero-bias conductance $G$ with the molecular size (length).
 
``Historically'',\cite{Beebe:06} the transition voltage $V_t$ was introduced as 
the bias at the minimum of the Fowler-Nordheim quantity 
$\log \left(I/V^2\right)$; the initial claim was that of a mechanistic 
transition from 
direct tunneling across a trapezoidal energy barrier ($V< V_t$) 
to Fowler-Nordheim (field-emission) tunneling across a triangular barrier
($V > V_t$).\cite{Beebe:06} 
As already noted in previous studies,\cite{Huisman:09,Araidai:10,Baldea:2010h} 
such a Fowler-Nordheim transition does \emph{not} occur in molecular junctions. 
\emph{Physically}, $V_t$ has \emph{nothing} to do with the Fowler-Nordheim tunneling theory.
\cite{Fowler:28,FowlerNordheim:28,Nordheim:28} In particular, in molecular junctions 
the quantity $\log \left(I/V^2\right)$ does \emph{not} linearly decrease 
with $1/V$ at higher biases; this was a result deduced
for extraction of electrons from cold metals by intense electric fields.\cite{FowlerNordheim:28}

Still, transport measurements in molecular junctions do yield 
curves for $\log(I/V^2)$ exhibiting a minimum. 
Mathematically, the bias $V=V_t$ at the minimum of $\log \left(I/V^2\right)$  
coincides with the bias where the differential conductance is two times large than 
the nominal conductance \cite{Baldea:2012e} % \cite{Baldea:2012b} 
%%%%%%%%%%%%%%%%%%%%%%%%%%%%%%%%%%%%%%%%%%%%%%%%%%%%%%%%%%%%
\begin{equation}
\label{eq-vt}
\left . \frac{\partial I}{\partial V}\right\vert_{V=V_t} = 2 \left . \frac{I}{V}\right\vert_{V=V_t}
\end{equation}
%%%%%%%%%%%%%%%%%%%%%%%%%%%%%%%%%%%%%%%%%%%%%%%%%%%%%%%%%%%%
Eqn~(\ref{eq-vt}) can be taken as alternative mathematical definition of $V_t$
revealing at the same time its physical meaning.
Using the minimum of $\log \left(I/V^2\right)$ plotted a function of $V$ (or $1/V$)\cite{Beebe:06} 
merely represents a mathematical trick to extract the $V_t$-value of eqn~(\ref{eq-vt}).
To eliminate the confusion that continues to exist in the literature 
on a mechanistic (Fowler-Nordheim) transition occurring at $V=V_t$, 
instead of Fowler-Nordheim diagrams $\log\left(I/V^2 \right)$ vs.~$V$, 
it might be more appropriate to use diagrams $V^2/I$ vs.~$V$, which have maxima exactly 
at the same $V=V_t$ given by eqn~(\ref{eq-vt});\cite{pvs} 
so, instead of ``transition voltage spectra'' (``TVS'', \figurename\ref{fig:generic-parabola}b), one can speak 
of a ``peak voltage spectra'' (``PVS'', \figurename\ref{fig:generic-parabola}c).
 
As expressed by eqn~(\ref{eq-vt}), 
$V_t$ represents a genuine \emph{nonequilibrium} property 
characterizing the charge transport through a molecular junction out of equilibrium. 
So, $V_t$ is \emph{qualitatively} different from the zero-bias conductance $G$;
in fact, the latter (as well as other properties commonly targeted in measurements, 
e.g., thermopower Seebeck coefficient) 
can be expressed by properties of a device (let it be molecular junction or else) 
at \emph{equilibrium} via fluctuation-dissipation theorem.\cite{Balescu:75}
Interestingly, as revealed by recent studies
\cite{Frisbie:11,Reddy:12a,Baldea:2012g} 
and supported by measurements comprising thousands of molecular junctions 
\cite{Guo:11,Tao:13},
it is even more justified to consider $V_t$ (rather than $G$) as a junction's characteristic property; 
in a given class of molecular junctions, $V_t$ is much less affected by inherent stochastic fluctuations
than individual $I-V$ traces or low bias conductances $G$.\cite{Guo:11} 
The aforementioned should be taken as motivation why $V_t$ represents a quantity on which 
the present transport study focuses.
%%%%%%%%%%%%%%%%%%%%%%%%%%%%%%%%%%%%%%%%%%%%%%%%%%%%%%%%%%%%%%%%%%%%%%%%%%%%%%%%%%%%%%%%%%%%%%%%%
\section{Models and relevant analytical formulas for symmetric $\mathbf{I-V}$ curves}
\label{sec:models}
%%%%%%%%%%%%%%%%%%%%%%%%%%%%%%%%%%%%%%%%%%%%%%%%%%%%%%%%%%%%%%%%%%%%%%%%%%%%%%%%%%%%%%%%%%%%%%%%% 
In this section we will examine a series of models widely employed in studies on charge 
transport through molecular devices. 
For some of these models, the dependence $I=I(V)$ or other useful formulas can be given 
in closed analytical form. Worthy to be remembered, describing idealized situations, 
these analytical results hold only if certain conditions are satisfied. Whenever the case, 
the conditions of validity will be also given below. 
Other models examined in this paper rely on tight-binding Hamiltonians, 
for which the exact dependence $I=I(V)$ can be obtained without substantial numerical effort.

For reasons presented later, the expansion coefficients $c_{2}$ and $c_{4}$
of the current up to the fifth order, eqn~(\ref{eq-i5}), will also be provided for each case
%%%%%%%%%%%%%%%%%%%%%%%%%%%%%%%%%%%%%%%%%%%%%%%%%%%%%%%%%%%%%%%%%%%%%%%%%%%%%%%%
\begin{equation}
I(V) = G V \left(1 + c_{2} V^2 + c_{4} V^4 \right) + \mathcal{O}\left(V^7\right) 
\label{eq-i5}
\end{equation}
%%%%%%%%%%%%%%%%%%%%%%%%%%%%%%%%%%%%%%%%%%%%%%%%%%%%%%%%%%%%%%%%%%%%%%%%%%%%%%%%%%%%%%%%%%%%%%%%%
They can be directly inserted in eqn~(\ref{eq-vt3}) and (\ref{eq-vt5}) to compute the transition voltages 
$V_{t3}$ and $V_{t 5}$ within the third and fifth order expansion, respectively
%%%%%%%%%%%%%%%%%%%%%%%%%%%%%%%%%%%%%%%%%%%%%%%%%%%%%%%%%%%%%%%%%%%%%%%%%%%%%%%%
\begin{eqnarray}
V_{t3} & = & \frac{1}{\sqrt{c_{2}}} 
\label{eq-vt3} \\
V_{t5} & = & 
\sqrt{
\frac{\sqrt{c_{2}^{2} + 12 c_{4}}-c_{2}}{6 c_{4}}}
\label{eq-vt5}
\end{eqnarray}
%%%%%%%%%%%%%%%%%%%%%%%%%%%%%%%%%%%%%%%%%%%%%%%%%%%%%%%%%%%%%%%%%%%%%%%%%%%%%%%%%%%%%%%%%%%%%%%%%
Notice that the expansion coefficients can be expressed as
%%%%%%%%%%%%%%%%%%%%%%%%%%%%%%%%%%%%%%%%%%%%%%%%%%%%%%%%%%%%%%%%%%%%%%%%%%%%%%%%
\begin{equation*}
c_{2} = 1/V_2^2 \ , \ c_{4} = 1/V_4^4
% \label{eq-coeff-35}
\end{equation*}
%%%%%%%%%%%%%%%%%%%%%%%%%%%%%%%%%%%%%%%%%%%%%%%%%%%%%%%%%%%%%%%%%%%%%%%%%%%%%%%%%%%%%%%%%%%%%%%%%
where $V_2$ and $V_4$ have dimensions of voltages and allows one to express the expansion
in terms of dimensionless voltages
%%%%%%%%%%%%%%%%%%%%%%%%%%%%%%%%%%%%%%%%%%%%%%%%%%%%%%%%%%%%%%%%%%%%%%%%%%%%%%%%
\begin{equation*}
I(V) = G V\left[1 + \left(V/V_2\right)^2 + \left(V/V_4\right)^4\right] + \mathcal{O}\left(V^7\right)
% \label{eq-i5-Vs}
\end{equation*}
%%%%%%%%%%%%%%%%%%%%%%%%%%%%%%%%%%%%%%%%%%%%%%%%%%%%%%%%%%%%%%%%%%%%%%%%%%%%%%%%%%%%%%%%%%%%%%%%%
%%%%%%%%%%%%%%%%%%%%%%%%%%%%%%%%%%%%%%%%%%%%%%%%%%%%%%%%%%%%%%%%%%%%%%%%%%%%%%%%

Next we briefly describe the models considered here and, whenever possible, 
give the relevant analytical formulas for the low bias conductance 
$G$ ($G_0 \equiv 2 e^2/h = 77.48\,\mu$S is the conductance quantum), the expansion coefficients 
$c_2$ and $c_4$ entering eqn~(\ref{eq-i5}) as well as 
the transition voltage $V_t$ along with the conditions for their validity.
 
%%%%%%%%%%%%%%%%%%%%%%%%%%%%%%%%%%%%%%%%%%%%%%%%%%%%%%%%%%%%%%%%%%%%%%%%%%%%%%%%%
(i) 
For biases $e \vert V \vert < \varepsilon_B$, within the 
Simmons WKB-based approximation for electron tunneling across a tunneling barrier of effective 
height $\varepsilon_B$ without lateral constriction
(Simmons's model), the current $I$ is given by 
eqn~(\ref{eq-i-simmons}).\cite{Simmons:63,Duke:69}  
The Taylor expansion of the RHS of eqn~(\ref{eq-i-simmons}) allows one to deduce formulas for the zero-bias conductance $G$
and the coefficients $c_2$ and $c_4$ entering eqn~(\ref{eq-i5}). 
They are expressed by eqn~(\ref{eq-G-simmons}), (\ref{eq-c3-simmons}), 
and (\ref{eq-c5-simmons}), respectively
%%%%%%%%%%%%%%%%%%%%%%%%%%%%%%%%%%%%%%%%%%%%%%%%%%%%%%%%%%%%%%%%%%%%%%%%%%%%%%%%%%%%%%%%%%%%%%%%% 
\begin{eqnarray}
I(V) & = & \frac{G}{e} \, 
\frac{2}{z - 2}
\left\{
\left(\varepsilon_B - \frac{e V}{2}\right) 
\exp\left[ 1 - z \sqrt{1 - \frac{e V}{2\varepsilon_B}}\right]
\right . \nonumber \\
& & 
\left . 
- \left(\varepsilon_B + \frac{e V}{2}\right) 
\exp\left[1 - z \sqrt{1 + \frac{e V}{2\varepsilon_B}}\right]
\right\} 
\label{eq-i-simmons} \\
G & = & G_0 \frac{\mathcal{A}}{d^2} \frac{z-2}{8 \pi}
\exp\left(-\tilde{\beta} d\right) , \ 
G \approx \mbox{constant } e^{-\tilde{\beta} d} 
\label{eq-G-simmons}  \\
\tilde{\beta} & \equiv & \alpha \sqrt{\varepsilon_B}
\label{eq-beta-tilde} \\
c_{2} & = & 0.0104167 \frac{z \left(z^2-3 z-3\right)}{(z-2) \varepsilon_B^2} 
\label{eq-c3-simmons} \\
c_{4} & = & 0.0000325521 \frac{ z\left(z^4-15 z^2-45 z-45\right)}{(z-2) \varepsilon_B^4} 
\label{eq-c5-simmons} 
%%%%%%%%%%%%%%%%%%%%%%
\end{eqnarray}
%%%%%%%%%%%%%%%%%%%%%%%%%%%%%%%%%%%%%%%%%%%%%%%%%%%%%%%%%%%%%%%%%%%%%%%%%%%%%%%%%%%%%%%%%%%%%%%%%
%%%%%%%%%%%%%%%%%%%%%%%%%%%%%%%%%%%%%%%%%%%%%%%%%%%%%%%%%%%%%%%%%%%%%%%%%%%%%%%%%%%%%%%%%%%%%%%%% 
\begin{equation}
\boxed{\mbox{Validity condition: } z \equiv \tilde{\beta} d  \agt  4 }
\label{eq-validity-simmons}
\end{equation} 
%%%%%%%%%%%%%%%%%%%%%%%%%%%%%%%%%%%%%%%%%%%%%%%%%%%%%%%%%%%%%%%%%%%%%%%%%%%%%%%%%%%%%%%%%%%%%%%%% 
Above, $\mathcal{A}$ stands for the junction's transverse area and 
%%%%%%%%%%%%%%%%%%%%%%%%%%%%%%%%%%%%%%%%%%%%%%%%%%%%%%%%%%%%%%%%%%%%%%%%%%%%%%%%%%%%%%%%%%%%%%%%% 
\begin{equation}
\label{eq-alpha}
\alpha \equiv 2\frac{\sqrt{2 m^\ast}}{\hbar} = \alpha_0 \sqrt{\frac{m^\ast}{m_0}} 
=  1.025\mbox{\,eV}^{-1}{\mbox{\AA}}^{-1/2} \sqrt{\frac{m^\ast}{m_0}} 
\end{equation} 
%%%%%%%%%%%%%%%%%%%%%%%%%%%%%%%%%%%%%%%%%%%%%%%%%%%%%%%%%%%%%%%%%%%%%%%%%%%%%%%%%%%%%%%%%%%%%%%%% 
$\alpha_0 \equiv 2 \sqrt{2 m_0}/{\hbar}$,
% = 1.025$\,eV$^{-1/2}{\mbox{\AA}}^{-1}$, 
$m_0$ and $m^\ast$ being the  Sommerfeld's constant \cite{Sommerfeld:33}, 
free and effective electron mass, respectively.
Eqn~(\ref{eq-validity-simmons}) (like eqn~(\ref{eq-validity-ib}) below) 
defines a dimensionless junction width $z$.
The transition voltage for the Simmons and Simmons-based models has been investigated in a series
of recent works.
% \cite{Huisman:09,Molen:11,Baldea:2012c,Baldea:2012e,Baldea:2012f,Baldea:2012h,Zandvliet:14,Baldea:2014d}
\cite{Huisman:09,Molen:11,Baldea:2012c,Baldea:2012e,Zandvliet:14,Baldea:2014d}

Eqn~(\ref{eq-i-simmons}) and (\ref{eq-G-simmons}) represent \emph{mathematical} approximate results; 
\ib{in addition
to the restriction on the bias range specified, they only hold for barriers satisfying 
eqn~(\ref{eq-validity-simmons}),
requiring tunneling barriers sufficiently high and wide.\cite{Gundlach:69b}}
In fact, the applicability of Simmons' approximation is more restrictive than 
required by eqn~(\ref{eq-validity-simmons}),
as revealed by a more detailed analysis.\cite{Duke:69,Miskovsky:82,Forbes:08}

%%%%%%%%%%%%%%%%%%%%%%%%%%%%%%%%%%%%%%%%%%%%%%%%%%%%%%%%%%%%%%%%%%%%%%%%%%%%%%%%%%%%%%%%%%%%%%%%%
(ii) For electron tunneling across a tunneling barrier with lateral constriction at biases $e \vert V \vert\alt \varepsilon_B$, 
the current $I$ can be expressed by eqn~(\ref{eq-i-ib}),\cite{Baldea:2012c} which applies provided that 
eqn~(\ref{eq-validity-ib}) is satisfied. In addition to the quantities $G$, $c_2$ and $c_4$ obtained by expanding
the RHS of eqn~(\ref{eq-i-ib}), the transition voltage $V_t$ can also be expressed in closed analytical form; 
see eqn~(\ref{eq-G-ib}), (\ref{eq-c35-ib}), and (\ref{eq-vt-ib})
%%%%%%%%%%%%%%%%%%%%%%%%%%%%%%%%%%%%%%%%%%%%%%%%%%%%%%%%%%%%%%%%%%%%%%%%%%%%%%%%%%%%%%%%%%%%%%%%% 
\begin{eqnarray}
I(V) & = & G \frac{4 \sqrt{\varepsilon_B}}{\alpha d} \sinh\left( \frac{\alpha d }{4 \sqrt{\varepsilon_B}} e V\right)  
\label{eq-i-ib} \\
% G & = & \underbrace{G_0 \frac{16}{\Phi} \frac{\tilde{\beta}^2}{\alpha^2}}_{G_{contact}} \exp\left(-\tilde{\beta} d\right) 
% G & = & G_{c}\exp\left(-\tilde{\beta} d\right) \ ; \ 
% G_{c} = G_0 \frac{16}{\Phi} \frac{\tilde{\beta}^2}{\alpha^2} \propto \frac{1}{\Phi} 
G & \propto & \exp\left(-\tilde{\beta} d\right)
\label{eq-G-ib} \\
c_{2} & = & \frac{z^2}{6\varepsilon_B^2} \ , \ c_{4} = \frac{z^4}{120\varepsilon_B^4} 
\label{eq-c35-ib} \\
V_t & = & 7.66003 \frac{\sqrt{\varepsilon_B}}{e\alpha d}
\label{eq-vt-ib}
\end{eqnarray}
%%%%%%%%%%%%%%%%%%%%%%%%%%%%%%%%%%%%%%%%%%%%%%%%%%%%%%%%%%%%%%%%%%%%%%%%%%%%%%%%%%%%%%%%%%%%%%%%
\begin{equation} 
\boxed{\mbox{ Validity condition: } z \equiv \tilde{\beta} d \agt 8 }
\label{eq-validity-ib}
\end{equation}
%%%%%%%%%%%%%%%%%%%%%%%%%%%%%%%%%%%%%%%%%%%%%%%%%%%%%%%%%%%%%%%%%%%%%%%%%%%%%%%%%%%%%%%%%%%%%%%%% 
Noteworthy, the restriction imposed by eqn~(\ref{eq-validity-ib})
to the results of eqn~(\ref{eq-i-ib}) and (\ref{eq-G-ib}) is more severe
than that expressed by eqn~(\ref{eq-validity-simmons}), which refers to a situation less appropriate
for molecular electronic devices. \ib{Like eqn~(\ref{eq-validity-simmons}), 
eqn~(\ref{eq-validity-ib}) expresses the fact that the above results for model (ii)
hold for physical situations where tunneling barriers are sufficiently high and wide.}

\begin{figure}[htb]
\centerline{\includegraphics[width=0.5\textwidth,angle=0]{fig_model.eps}\hspace*{12ex}}
$ $\\[-12ex]
\centerline{\includegraphics[width=0.4\textwidth,angle=0]{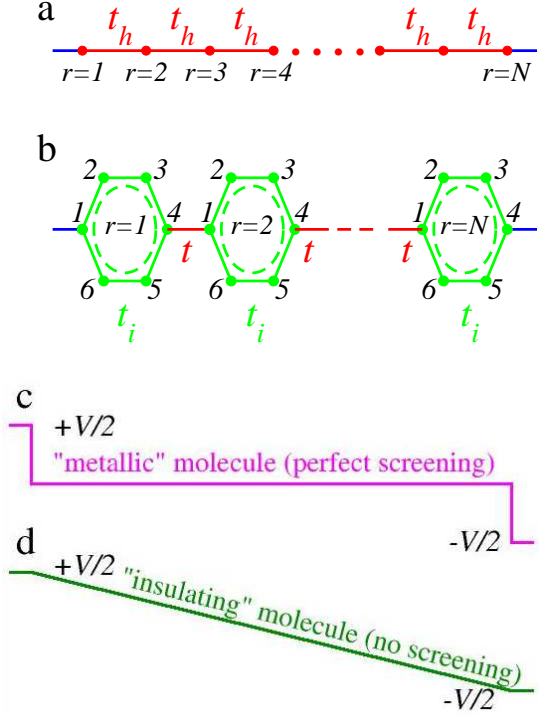}}
% $ $\\[0ex]
\caption{Panels (a) and (b) represent schematically the tight binding models described by 
% eqn~(\ref{eq-ham-chains}) and (\ref{eq-ham-rings}).
eqn~\ib{(S1)} and \ib{(S3)}.
Panels (c) and (d) depict the cases of a constant potential and a linear potential drop, respectively, 
as limiting cases of spatial potential profiles between electrodes under a bias $V$.}
\label{fig:tba-models}
\end{figure}

(iii) The highly off-resonant sequential tunneling 
\cite{Mujica:01} (superexchange limit \cite{McConnell:61})
across a molecular bridge consisting of a wire
with $N$ sites and one energy level ($\varepsilon_B$) per site
represents a limiting case of the situation depicted in \figurename\ref{fig:tba-models}a. 
Provided that eqn~(\ref{eq-validity-mr}) is satisfied, the current is expressed by eqn~(\ref{eq-i-mr});
eqn~(\ref{eq-G-mr}) and (\ref{eq-c35-mr}) follow via straightforward series expansion.
%%%%%%%%%%%%%%%%%%%%%%%%%%%%%%%%%%%%%%%%%%%%%%%%%%%%%%%%%%%%%%%%%%%%%%%%%%%%%%%%%%%%%%%%%%%%%%%%% 
\begin{eqnarray}
I(V) & = & \frac{G}{e}\,\frac{\varepsilon_B}{2 N - 1} \times \nonumber \\
     & & \left[
\left(1 - \frac{e V}{2\varepsilon_B}\right)^{1 - 2 N} -
\left(1 + \frac{e V}{2\varepsilon_B}\right)^{1 - 2 N} 
\right] 
\label{eq-i-mr} \\
% \end{eqnarray}
%%%%%%%%%%%%%%%%%%%%%%%%%%%%%%%%%%%%%%%%%%%%%%%%%%%%%%%%%%%%%%%%%%%%%%%%%%%%%%%%%%%%%%%%%%%%%%%%% 
%%%%%%%%%%%%%%%%%%%%%%%%%%%%%%%%%%%%%%%%%%%%%%%%%%%%%%%%%%%%%%%%%%%%%%%%%%%%%%%%%%%%%%%%%%%%%%%%% 
% \begin{eqnarray}
G & = & G_0 \frac{\Gamma^2}{t_{h}^2}\,\frac{t_{h}^{2 N}}{\varepsilon_B^{2 N}}, 
\ G \propto e^{- \beta N}, \ \beta = 2 \log \frac{\varepsilon_B}{t_h} 
\label{eq-G-mr} \\
c_{2} & = & \frac{N (2N + 1)}{12 \varepsilon_B^2} \nonumber \\
c_{4} & = & \frac{N ( N+1) (2N+1) (2N+3)}{480 \varepsilon_B^4} 
\label{eq-c35-mr} 
\end{eqnarray}

%%%%%%%%%%%%%%%%%%%%%%%%%%%%%%%%%%%%%%%%%%%%%%%%%%%%%%%%%%%%%%%%%%%%%%%%%%%%%%%%%%%%%%%%%%%%%%%%% 
\begin{equation}
\boxed{\mbox{Validity condition: } \varepsilon_B \gg  t_h, \Gamma }
\label{eq-validity-mr} 
\end{equation}
%%%%%%%%%%%%%%%%%%%%%%%%%%%%%%%%%%%%%%%%%%%%%%%%%%%%%%%%%%%%%%%%%%%%%%%%%%%%%%%%%%%%%%%%%%%%%%%%% 
Here, $t_h$ is the hopping integral between nearest neighboring sites. 
Eqn~(\ref{eq-i-mr}) and (\ref{eq-G-mr}) emerge 
% straightforwardly 
from the Landauer
formula, eqn~(\ref{eq-landauer}), using the transmission given in the RHS of 
% eqn~(\ref{eq-T-2-sites})
eqn~\ib{(S2)}
for arbitrary $N$. Notice that the power dependent transmission 
expressed by the latter yields a strict proportionality \cite{typo}
%%%%%%%%%%%%%%%%%%%%%%%%%%%%%%%%%%%%%%%%%%%%%%%%%%%%%%%%%%%%%%%%%%%%%%%%%%%%%%%%%%%%%%%%%%%%%%%%% 
\begin{equation}
\last{e} V_t = \last{2} u_t \varepsilon_B \propto \varepsilon_B
\label{eq-Vt-mr} 
\end{equation}
%%%%%%%%%%%%%%%%%%%%%%%%%%%%%%%%%%%%%%%%%%%%%%%%%%%%%%%%%%%%%%%%%%%%%%%%%%%%%%%%%%%%%%%%%%%%%%%%% 
where the dimensionless quantity $u_t$ is the solution of the parameter-free algebraic 
eqn~(\ref{eq-ut-mr}), which results by inserting eqn~(\ref{eq-i-mr}) into eqn~(\ref{eq-vt})
%%%%%%%%%%%%%%%%%%%%%%%%%%%%%%%%%%%%%%%%%%%%%%%%%%%%%%%%%%%%%%%%%%%%%%%%%%%%%%%%%%%%%%%%%%%%%%%%% 
\begin{eqnarray}
& & \left(N - 1/2\right) u_t \left[ \left(1 + u_t\right)^{2 N} + \left(1 - u_t\right)^{2 N}\right] \nonumber \\
& & = \left(1 - u_t\right) \left(1 + u_t\right)^{2 N} - \left(1 + u_t\right) \left(1 - u_t\right)^{2 N}
\label{eq-ut-mr} 
\end{eqnarray}
%%%%%%%%%%%%%%%%%%%%%%%%%%%%%%%%%%%%%%%%%%%%%%%%%%%%%%%%%%%%%%%%%%%%%%%%%%%%%%%%%%%%%%%%%%%%%%%%% 
\ib{Eqn~(\ref{eq-validity-mr}) expresses the physical fact that the above results for model (iii) hold 
for situations corresponding to strongly off-resonant tunneling.} 
%%%%%%%%%%%%%%%%%%%%%%%%%%%%%%%%%%%%%%%%%%%%%%%%%%%%%%%%%%%%%%%%%%%%%%%%%%%%%%%%%%%%%%%%%%%%%%%%% 

(iv) A molecular chain consisting of $N$ monomers, each monomer being characterized by a
single orbital
%EIS single orbital, can be described by the following second-quantized tight-binding Hamiltonian 
%%%%%%%%%%%%%%%%%%%%%%%%%%%%%%%%%%%%%%%%%%%%%%%%%%%%%%%%%%%%%%%%%%%%%%%%%%%%%%%%%%%%%%%%%%%%%%%%% 
%ESI \begin{equation}
%ESI \label{eq-ham-chains}
%ESI H_1 = \sum_{r=1}^{N} \underbrace{\left(\varepsilon_B - e V_r\right)}_{\varepsilon_B^r} c_{r}^{\dagger} c_{r} 
%ESI - t_h  \sum_{r=1}^{N-1} \left( c_{r}^{\dagger} c_{r+1} + c_{r+1}^{\dagger} c_{r} \right) 
%ESI \end{equation}
%%%%%%%%%%%%%%%%%%%%%%%%%%%%%%%%%%%%%%%%%%%%%%%%%%%%%%%%%%%%%%%%%%%%%%%%%%%%%%%%%%%%%%%%%%%%%%%%%
%ESI where $c^\dagger$ and $c$ are creation and annihilation operators for electrons whose spin is 
%ESI omitted for simplicity. 
%ESI It is schematically depicted in \figurename\ref{fig:tba-models}a. 
is schematically depicted in \figurename\ref{fig:tba-models}a.
Model (iii) presented above represents a limiting case 
(cf.~eqn~(\ref{eq-validity-mr})) of this model. 
To illustrate this, we give in 
% eqn (\ref{eq-T-2-sites}) 
eqn~\ib{(S2)} of the {\si} 
the transmission function computed exactly and within the sequential tunneling approximation for the case $N=2$
%ESI 
%ESI To illustrate this, we give 
%ESI below the transmission function computed exactly and within the sequential tunneling approximation for the case $N=2$
%%%%%%%%%%%%%%%%%%%%%%%%%%%%%%%%%%%%%%%%%%%%%%%%%%%%%%%%%%%%%%%%%%%%%%%%%%%%%%
%ESI \begin{eqnarray}
%ESI & & T_{\mbox{exact}}(\varepsilon) = 
%ESI \frac{\Gamma^2 t_h^2}
%ESI {\left[\left(\varepsilon - \varepsilon_B\right)^2 - t_h^2 - \Gamma^2/4\right]^2 
%ESI + \varepsilon_B^2 \Gamma^2} \nonumber \\
%ESI & & \xlongrightarrow[\text{}]{\text{$t_h, \Gamma \ll \varepsilon_B$}}
%ESI T_{\mbox{superexchange}}(\varepsilon) = \left . \frac{\Gamma^2 t_h^{2(N-1)}}{\left(\varepsilon - \varepsilon_B\right)^{2N}}\right\vert_{N=2} 
%ESI \label{eq-T-2-sites}
%ESI \end{eqnarray}
%%%%%%%%%%%%%%%%%%%%%%%%%%%%%%%%%%%%%%%%%%%%%%%%%%%%%%%%%%%%%%%%%%%%%%%%%%%%%%
Adapted to oligophenylene chains, 
$\varepsilon_B$ should be taken as a model for the HOMO energy of an isolated phenylene unit, and $t_h$ 
as the coupling between HOMO levels of adjacent rings. (This is the motivation
for choosing the subscript $h$.)

Under applied bias $V$, two limiting cases can be considered. One limit is that of an applied field 
completely screened out by delocalized electrons (``metallic'' molecule, \figurename\ref{fig:tba-models}c);
the potential is constant across the molecule ($V_r = 0$), and the site energies $\varepsilon_B^r$ 
are the same $\varepsilon_B^r \equiv \varepsilon_B$.
Another limit (no screening, ``insulating'' molecule) is that of a potential 
$V_r$ varying linearly from site to site between the 
values $+V/2$ and $-V/2$ at the two electrodes 
(\figurename\ref{fig:tba-models}d), yielding 
$r$-dependent on-site energies $\varepsilon_B^r = \varepsilon_B - e V_r$. 
 
%%%%%%%%%%%%%%%%%%%%%%%%%%%%%%%%%%%%%%%%%%%%%%%%%%%%%%%%%%%%%%%%%%%%%%%%%%%%%%%%%%%%%%%%%%%%%%%%% 
(v) Molecular wires consisting of six-site rings with a single ($\pi$-electron) energy level per 
site (CH-unit) depicted in \figurename\ref{fig:tba-models}b represent a tight-binding description of 
oligophenylene molecules. The corresponding second-quantized Hamiltonian is given in the {\si}.
%ESI The corresponding second-quantized Hamiltonian reads
%%%%%%%%%%%%%%%%%%%%%%%%%%%%%%%%%%%%%%%%%%%%%%%%%%%%%%%%%%%%%%%%%%%%%%%%%%%%%%%%%%%%%%%%%%%%%%%%% 
%ESI \begin{eqnarray}
%ESI H_2 & = & \varepsilon_B \sum_{r=1}^{N} \sum_{j=1}^{6}c_{j, r}^{\dagger} c_{j, r} 
%ESI  - t_i  \sum_{r=1}^{N-1} \sum_{j=1}^{6} \left( c_{j, r}^{\dagger} c_{j, r+1} + c_{j, r+1}^{\dagger} c_{j, r} \right) \nonumber \\
%ESI & & - t  \left( c_{4, 1}^{\dagger} c_{1, 2} + c_{1, 2}^{\dagger} c_{4, 1} + 
%ESI              c_{4, 2}^{\dagger} c_{1, 3} + c_{1, 3}^{\dagger} c_{4, 2} + \ldots \right)
%ESI \label{eq-ham-rings}
%ESI \end{eqnarray}
%%%%%%%%%%%%%%%%%%%%%%%%%%%%%%%%%%%%%%%%%%%%%%%%%%%%%%%%%%%%%%%%%%%%%%%%%%%%%%%%%%%%%%%%%%%%%%%%% 
As illustrated by \figurename\ref{fig:tba-models}b, $t_i$ and $t$ stand for intra- and inter-ring 
hopping integrals, respectively.

%%%%%%%%%%%%%%%%%%%%%%%%%%%%%%%%%%%%%%%%%%%%%%%%%%%%%%%%%%%%%%%%%%%%%%%%%%%%%%%%%%%%%%%%%%%%%%%%%%  
%%%%%%%%%%%%%%%%%%%%%%%%%%%%%%%%%%%%%%%%%%%%%%%%%%%%%%%%%%%%%%%%%%%%%%%%%%%%%%%%%%%%%%%%%%%%%%%%%%  

Although for models (iv) and (v) the dependence $I=I(V)$ cannot be expressed in closed analytical form,
it can easily be obtained numerically within the Landauer framework \cite{HaugJauho}
%%%%%%%%%%%%%%%%%%%%%%%%%%%%%%%%%%%%%%%%%%%%%%%%%%%%%%%%%%%%%%%%%%%%%%%%%%%%%%%%%%%%%%%%%%%%%%%%%
\begin{equation}
\label{eq-landauer}
I(V) = \frac{G_0}{e} \int_{-eV/2}^{eV/2} T(E)\,d\,E 
\end{equation}
%%%%%%%%%%%%%%%%%%%%%%%%%%%%%%%%%%%%%%%%%%%%%%%%%%%%%%%%%%%%%%%%%%%%%%%%%%%%%%%%%%%%%%%%%%%%%%%%%
Analytical formulas for the transmission function $T(E)$ are lengthy even for
$\Gamma$ much smaller than $\varepsilon_B$ and $t_h$'s,\cite{Onipko:98,Schmickler:11} 
so they are of little interest
in the present context. The numerical effort to compute 
$T(E)$ via the Landauer trace formula is insignificant: it consists of a numerical integration
and a matrix inversion, which is needed to compute the retarded Green's function.\cite{HaugJauho} 

%%%%%%%%%%%%%%%%%%%%%%%%%%%%%%%%%%%%%%%%%%%%%%%%%%%%%%%%%%%%%%%%%%%%%%%%%%%%%%%%%%%%%%%%%%%%%%%%%
(vi) We will also examine the case of a generalized exponential transmission
%%%%%%%%%%%%%%%%%%%%%%%%%%%%%%%%%%%%%%%%%%%%%%%%%%%%%%%%%%%%%%%%%%%%%%%%%%%%%%%%%%%%%%%%%%%%%%%%%
\begin{equation}
\label{eq-T-delta}
T_{\delta}(E) = \exp\left[ - \frac{\left\vert E - \varepsilon_B\right\vert^\delta}{\Delta^\delta}\right] 
\end{equation}
%%%%%%%%%%%%%%%%%%%%%%%%%%%%%%%%%%%%%%%%%%%%%%%%%%%%%%%%%%%%%%%%%%%%%%%%%%%%%%%%%%%%%%%%%%%%%%%%%
Results for this case, including 
an approximate analytical formula for the current valid for biases not considerably exceeding $V_t$ 
and $\varepsilon_B \gg \Delta$ that can be deduced resorting to a 
Stratton-like approximation 
% \cite{Stratton:62a,Baldea:2012f}
\cite{Stratton:62a,Baldea:2012e}
for transmission are presented below. The conductance $G$ of eqn~(\ref{eq-G-delta}) 
and the coefficients $c_{2,4}$ of eqn~(\ref{eq-c2-delta}) and (\ref{eq-c4-delta}) 
can be straightforwardly obtained from the fifth-order expansion of the RHS of 
eqn~(\ref{eq-i-delta-exact})
%%%%%%%%%%%%%%%%%%%%%%%%%%%%%%%%%%%%%%%%%%%%%%%%%%%%%%%%%%%%%%%%%%%%%%%%%%%%%%%%%%%%%%%%%%%%%%%%% 
\begin{eqnarray}
I(V) & = & \frac{G_0}{e} \int_{-eV/2}^{eV/2} T_\delta(E)\,d\,E 
\label{eq-i-delta-exact}\\
I(V) & {\raisebox{-0.3ex}{$\stackrel{\Delta \ll \varepsilon_{B}}{\approx}$}} & 
I^{approx} (V) \nonumber \\
& \equiv &
G\,\frac{2\varepsilon_B}{e\delta} \left(\frac{\Delta}{\varepsilon_B}\right)^\delta 
\sinh\left[\delta\left(\frac{\varepsilon_B}{\Delta}\right)^\delta\frac{e V}{2\varepsilon_B}\right] 
\label{eq-i-delta-approx} \\
G & = & G_0 \exp\left(-\frac{\varepsilon_B^\delta}{\Delta^\delta}\right) 
\label{eq-G-delta} \\
V_{t}^{approx} 
& {\raisebox{-0.3ex}{$\stackrel{\Delta \ll \varepsilon_{B}}{\approx}$}} & 
1.91501 \frac{2\varepsilon_B}{\delta e}
\left(\frac{\Delta}{\varepsilon_B}\right)^\delta  
\label{eq-vt-approx-delta} 
\end{eqnarray}
%%%%%%%%%%%%%%%%%%%%%%%%%%%%%%%%%%%%%%%%%%%%%%%%%%%%%%%%%%%%%%%%%%%%%%%%%%%%%%%%%%%%%%%%%%%%%%%%% 
\begin{eqnarray}
c_{2} & = & 
\frac{e^2 \delta}{24\varepsilon_B^2}
\left(\frac{\varepsilon_B}{\Delta}\right)^\delta
\left[
\left( \frac{\varepsilon_B}{\Delta} \right)^\delta \delta
- \delta + 1
\right] \nonumber \\
& {\raisebox{-0.3ex}{$\stackrel{\Delta \ll \varepsilon_{B}}{\approx}$}} &
c_{2}^{approx} \equiv 
\frac{ e^2 \delta^2 }{24\varepsilon_B^2}\left(\frac{\varepsilon_B}{\Delta}\right)^{2\delta} 
\label{eq-c2-delta} \\
c_{4} & = & 
\frac{e^4 \delta}{1920\varepsilon_B^4}
\left(\frac{\varepsilon_B}{\Delta}\right)^\delta 
 \left\{
\left(\frac{\varepsilon_B}{\Delta }\right)^{3 \delta }   \delta ^3 
  - \left(\delta - 1\right) \times \right . \nonumber \\
& &  \left . \left[  
    6 \delta ^2 \left(\frac{\varepsilon_B}{\Delta }\right)^{2 \delta }
    - \delta \left(7\delta - 11\right)  \left(\frac{\varepsilon_B}{\Delta }\right)^{\delta } + \right . \right . \nonumber \\
    & & \left . \left . \left(\delta - 2\right) \left(\delta - 3\right) 
    \right]
\right\}  \label{eq-c4-delta} \\
& 
{\raisebox{-0.3ex}{$\stackrel{\Delta \ll \varepsilon_{B}}{\approx}$}} 
& c_{4}^{approx} \equiv \frac{\delta^4 e^4}{1920\varepsilon_B^4} \left(\frac{\varepsilon_B}{\Delta}\right)^{4\delta} 
{\raisebox{-0.3ex}{$\stackrel{\Delta \ll \varepsilon_{B}}{\approx}$}} 
\frac{3}{10} c_{2}^2 
\label{eq-c4-approx-delta}
\end{eqnarray}
The particular case $\delta=2$, which corresponds to a Gaussian transmission, 
has occasionally been studied in the literature \cite{Araidai:10,Vilan:13}; results for 
this case are presented in the {\si}.

The relation given below holds in general (not only for the Gaussian transmission $\delta \equiv 2$)
%%%%%%%%%%%%%%%%%%%%%%%%%%%%%%%%%%%%%%%%%%%%%%%%%%%%%%%%%%%%%%%%%%%%%%%%%%%%%%%%%%%%%%%%%%%%%%%%%
\begin{equation}
\label{eq-vt0-vt3-vt5-gauss-general}
V_{t,5}^{approx} = 0.797483 V_{t,3}^{approx} = 1.02006 V_{t}^{approx} 
\end{equation}
%%%%%%%%%%%%%%%%%%%%%%%%%%%%%%%%%%%%%%%%%%%%%%%%%%%%%%%%%%%%%%%%%%%%%%%%%%%%%%%%%%%%%%%%%%%%%%%%%
Here, $V_{t,3}^{approx}$ and $V_{t,5}^{approx}$ represent estimates obtained from 
eqn~(\ref{eq-vt3}) and (\ref{eq-vt5}) using the approximate coefficients 
$c_{2}^{approx}$ and $c_{4}^{approx}$ given above instead of $c_{2}$ and $c_{4}$, respectively.
An expression similar to $V_{t,3}^{approx}$ has been given previously.\cite{Vilan:13}
\ib{Notice that above the superscript \emph{approx} refers to physical 
situations of strongly off-resonant tunneling,
mathematically expressed by the inequality $\Delta \ll \varepsilon_B$.}
%%%%%%%%%%%%%%%%%%%%%%%%%%%%%%%%%%%%%%%%%%%%%%%%%%%%%%%%%%%%%%%%%%%%%%%%%%%%%%%%%%%%%%%%%%%%%%%%% 
%%%%%%%%%%%%%%%%%%%%%%%%%%%%%%%%%%%%%%%%%%%%%%%%%%%%%%%%%%%%%%%%%%%%%%%%%%%%%%%%%%%%%%%%%%%%%%%%% 
\section{Results for symmetric $\mathbf{I-V}$ curves using generic model parameter values}
\begin{figure*}[htb]
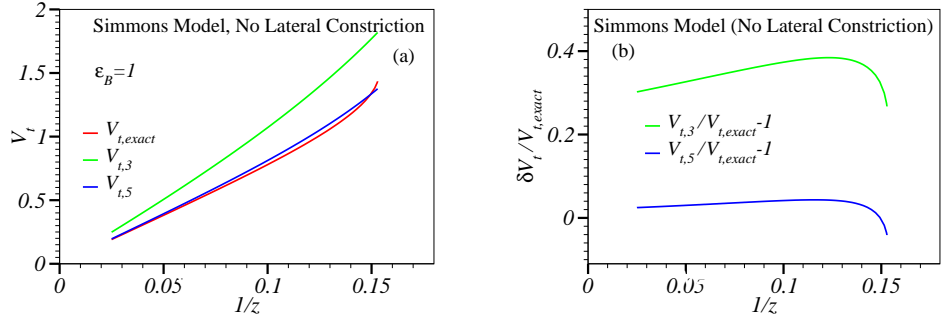

\centerline{\includegraphics[width=0.32\textwidth,angle=0]{fig_vt_simmons.eps}
\hspace*{5ex}
\includegraphics[width=0.32\textwidth,angle=0]{fig_vt_relativ_simmons.eps}}
% $ $\\[0ex]
\caption{The transition voltage $V_{t,exact}$ computed by using the ``exact'' 
current for the Simmons model [termed model (i) in the main text], 
eqn~(\ref{eq-i-simmons}) (panel (a)), along with the approximate estimates $V_{t,3}$ and $V_{t,5}$ obtained by 
inserting the expansion coefficients $c_{2}$ and $c_{4}$ of eqn~(\ref{eq-c3-simmons}) and (\ref{eq-c5-simmons}) 
into eqn~(\ref{eq-vt3}) and (\ref{eq-vt5}). Panel (b) show that, while $V_{t,3}$ significantly
deviates from $V_{t,exact}$ (typical experimental uncertainties in $V_t$ do not exceed 
$\sim 10$\% \cite{Tao:13,baldea:2015}), $V_{t,5}$ agrees with $V_{t,exact}$ within a few percents 
($z \equiv \tilde{\beta}d$ is a dimensionless barrier width).}
\label{fig:simmons}
\end{figure*}

\begin{figure*}[htb]
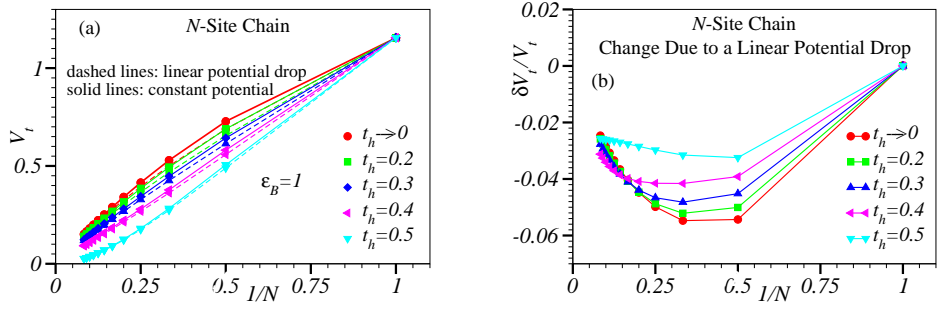

\centerline{\includegraphics[width=0.32\textwidth,angle=0]{fig_vt_mr_potenial_profile.eps}
\hspace*{5ex}
\includegraphics[width=0.32\textwidth,angle=0]{fig_vt_mr_relative_change_due_to_potenial_profile.eps}}
% $ $\\[0ex]
\caption{Transition voltage $V_t$ computed exactly for a molecular wire whose monomers are modeled 
as single sites characterized by a single orbital of energy $\varepsilon_B$. For arbitrary 
nearest-neighbor hopping integral values $t_h$, this is termed model (iv) in the main text, which becomes 
model (iii) in the limit $t_h \ll \varepsilon_B$. Notice the almost insignificant change  
$V_t \to V_t + \delta V_t$ when the spatial potential profile across the chain changes 
from constant (\figurename\ref{fig:tba-models}c) to a linear drop (\figurename\ref{fig:tba-models}d). 
The differences between the dashed and solid lines in 
panel (a) represent the absolute changes $\delta V_t$. 
The relative changes $\delta V_t/V_t$ are shown in panel (b).}
\label{fig:mr-V-profile}
\end{figure*}

%%%%%%%%%%%%%%%%%%%%%%%%%%%%%%%%%%%%%%%%%%%%%%%%%%%%%%%%%%%%%%%%%%%%%%%%%%%%%%%%%%%%%%%%%%%%%%%%% 
\label{sec:generic-parameters}
%%%%%%%%%%%%%%%%%%%%%%%%%%%%%%%%%%%%%%%%%%%%%%%%%%%%%%%%%%%%%%%%%%%%%%%%%%%%%%%%%%%%%%%%%%%%%%%%%
Detailed results for current ($I$) and transition voltages computed exactly 
and within the third- and fifth-order expansions 
($V_{t,exact} \equiv V_t$, $V_{t,3}$, and $V_{t,5}$, respectively) are collected in 
\figurename\ref{fig:simmons}, 
\ib{S1}, % \ref{fig:sinh}, 
\ref{fig:mr}, 
\ref{fig:mr-V-profile}, 
\ib{S2}, % \ref{fig:n=2}, 
\ib{S3}, % \ref{fig:n=3}, 
\ib{S4}, % \ref{fig:n=4}, 
\ref{fig:ndr}, 
\ref{fig:gaussian-eB}, 
\ib{S5} % \ref{fig:gaussian-sigma}
and \ref{fig:iv-fn-gaussian}.
Below we will emphasize some main aspects of the results obtained for 
the corresponding numerical simulations done with generic parameter values.  
%%%%%%%%%%%%%%%%%%%%%%%%%%%%%%%%%%%%%%%%%%%%%%%%%%%%%%%%%%%%%%%%%%%%%%%%%%%%%%%%%%%%%%%%%%%%%%%%% 
\subsection{The need to go beyond the parabolic conductance approximation}
\begin{figure*}[htb]
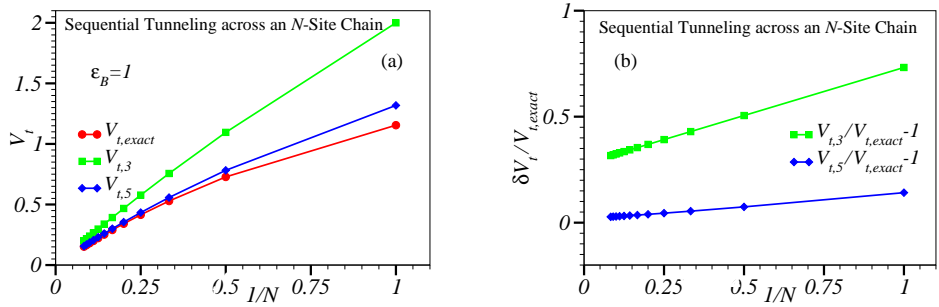

\centerline{\includegraphics[width=0.32\textwidth,angle=0]{fig_vt_mr.eps}
\hspace*{5ex}
\includegraphics[width=0.32\textwidth,angle=0]{fig_vt_relativ_mr.eps}}
% $ $\\[0ex]
\caption{(a) The transition voltage $V_{t,exact}$ computed from the
current of eqn~(\ref{eq-i-mr}) corresponding to the superexchange mechanism across a molecular wire 
modeled as chain of sites having a single orbital of 
energy $\varepsilon_B$ [termed model (iii) above], 
along with the approximate estimates $V_{t,3}$ and $V_{t,5}$ obtained by 
inserting the expansion coefficients $c_{2}$ and $c_{4}$ of eqn~(\ref{eq-c35-mr}) 
into eqn~(\ref{eq-vt3}) and (\ref{eq-vt5}) (panel (a)). 
Panel (b) shows that $V_{t,3}$ 
deviates from $V_{t,exact}$ by up to 73\% (much larger than typical experimental uncertainties in $V_t$ of 
$\sim 10$\% \cite{Tao:13,baldea:2015}), while $V_{t,5}$ agrees with $V_{t,exact}$ within 
at most 14\%.
Notice that, within the sequential tunneling approximation,
$V_{t,exact}$, $V_{t,3}$, and $V_{t,5}$ do not depend on the inter-site hopping integral $t_h$.}
\label{fig:mr}
\end{figure*}

%%%%%%%%%%%%%%%%%%%%%%%%%%%%%%%%%%%%%%%%%%%%%%%%%%%%%%%%%%%%%%%%%%%%%%%%%%%%%%%%%%%%%%%%%%%%%%%%% 
\label{sec:beyond-parabolic}
%%%%%%%%%%%%%%%%%%%%%%%%%%%%%%%%%%%%%%%%%%%%%%%%%%%%%%%%%%%%%%%%%%%%%%%%%%%%%%%%%%%%%%%%%%%%%%%%%
As visible in the aforementioned figures, the estimate $V_{t,5}$, based on the fifth-order 
expansion of eqn~(\ref{eq-i5}), represents a good approximation for the transition voltage 
$V_{t,exact} \equiv V_t$ computed exactly for each model considered. For reasonably broad 
ranges of the model parameter values, the relative deviations 
$\left\vert V_{t,5} - V_{t}\right\vert/V_t$ 
usually fall within typical experimental errors ($\alt 10$\%).\cite{Tao:13,baldea:2015} 
This does \emph{not} apply to the (over)estimate $V_{t,3}$, whose deviations from the exact 
values are considerably larger that experimental errors. As visible in 
\figurename\ref{fig:mr} and 
\ib{S8}, % \figurename\ref{fig:vt3-vt5-chains-rings-opds}, 
$V_{t,3}$
could be almost two times larger that $V_{t,exact}$. 
This demonstrates that the cubic approximation for current (or parabolic approximation 
for conductance), eqn~(\ref{eq-i3}), can only be used for qualitative purposes.
%%%%%%%%%%%%%%%%%%%%%%%%%%%%%%%%%%%%%%%%%%%%%%%%%%%%%%%%%%%%%%%%%%%%%%%%%%%%%%%%%%%%%%%%%%%%%%%%%%  
%%%%%%%%%%%%%%%%%%%%%%%%%%%%%%%%%%%%%%%%%%%%%%%%%%%%%%%%%%%%%%%%%%%%%%%%%%%%%%%%%%%%%%%%%%%%%%%%%%  
%%%%%%%%%%%%%%%%%%%%%%%%%%%%%%%%%%%%%%%%%%%%%%%%%%%%%%%%%%%%%%%%%%%%%%%%%%%%%%%%%%%%%%%%%%%%%%%%%%  
%ESI \begin{figure*}[htb]
% 
%ESI \centerline{\includegraphics[width=0.32\textwidth,angle=0]{fig_vt_sinh.eps}\hspace*{5ex}
% \includegraphics[width=0.32\textwidth,angle=0]{fig_vt_relativ_sinh.eps}}
% $ $\\[0ex]
%ESI \caption{(a) The transition voltage $V_{t,exact}$ computed from the
%ESI current of eqn~(\ref{eq-i-ib}) due to laterally constricted electrons tunneling across an energy barrier 
%ESI [termed model (ii) above], 
%ESI along with the approximate estimates $V_{t,3}$ and $V_{t,5}$ obtained by 
%ESI inserting the expansion coefficients $c_{2}$ and $c_{4}$ of eqn~(\ref{eq-c35-ib}) 
%ESI into eqn~(\ref{eq-vt3}) and (\ref{eq-vt5}) (panel (a). 
%ESI Panel (b) shows that $V_{t,3}$ 
%ESI deviates from $V_{t,exact}$ by 28\% (larger than typical experimental uncertainties in $V_t$ of 
%ESI $\sim 10$\% \cite{Tao:13,baldea:2015}), while $V_{t,5}$ agrees with $V_{t,exact}$ within 2\%.
%ESI Notice that the products $a V_t$, $a V_{t,3}$, and $a V_{t,5}$ do not depend on the parameter 
%ESI $a \equiv \alpha d e/\left(4\sqrt{\varepsilon_B}\right)$.}
%ESI \label{fig:sinh}
%ESI \end{figure*}
%%%%%%%%%%%%%%%%%%%%%%%%%%%%%%%%%%%%%%%%%%%%%%%%%%%%%%%%%%%%%%%%%%%%%%%%%%%%%%%%%%%%%%%%%%%%%%%%%%  
%%%%%%%%%%%%%%%%%%%%%%%%%%%%%%%%%%%%%%%%%%%%%%%%%%%%%%%%%%%%%%%%%%%%%%%%%%%%%%%%%%%%%%%%%%%%%%%%%  
%%%%%%%%%%%%%%%%%%%%%%%%%%%%%%%%%%%%%%%%%%%%%%%%%%%%%%%%%%%%%%%%%%%%%%%%%%%%%%%%%%%%%%%%%%%%%%%%%%  
%%%%%%%%%%%%%%%%%%%%%%%%%%%%%%%%%%%%%%%%%%%%%%%%%%%%%%%%%%%%%%%%%%%%%%%%%%%%%%%%%%%%%%%%%%%%%%%%% 
\subsection{The spatial potential profile as possible source of negative differential resistance}
%%%%%%%%%%%%%%%%%%%%%%%%%%%%%%%%%%%%%%%%%%%%%%%%%%%%%%%%%%%%%%%%%%%%%%%%%%%%%%%%%%%%%%%%%%%%%%%%% 
\label{sec:ndr}
%%%%%%%%%%%%%%%%%%%%%%%%%%%%%%%%%%%%%%%%%%%%%%%%%%%%%%%%%%%%%%%%%%%%%%%%%%%%%%%%%%%%%%%%%%%%%%%%%
%%%%%%%%%%%%%%%%%%%%%%%%%%%%%%%%%%%%%%%%%%%%%%%%%%%%%%%%%%%%%%%%%%%%%%%%%%%%%%%%%%%%%%%%%%%%%%%%% 
%%%%%%%%%%%%%%%%%%%%%%%%%%%%%%%%%%%%%%%%%%%%%%%%%%%%%%%%%%%%%%%%%%%%%%%%%%%%%%%%%%%%%%%%%%%%%%%%%
%%%%%%%%%%%%%%%%%%%%%%%%%%%%%%%%%%%%%%%%%%%%%%%%%%%%%%%%%%%%%%%%%%%%%%%%%%%%%%%%%%%%%%%%%%%%%%%%% 
\begin{figure*}[htb]
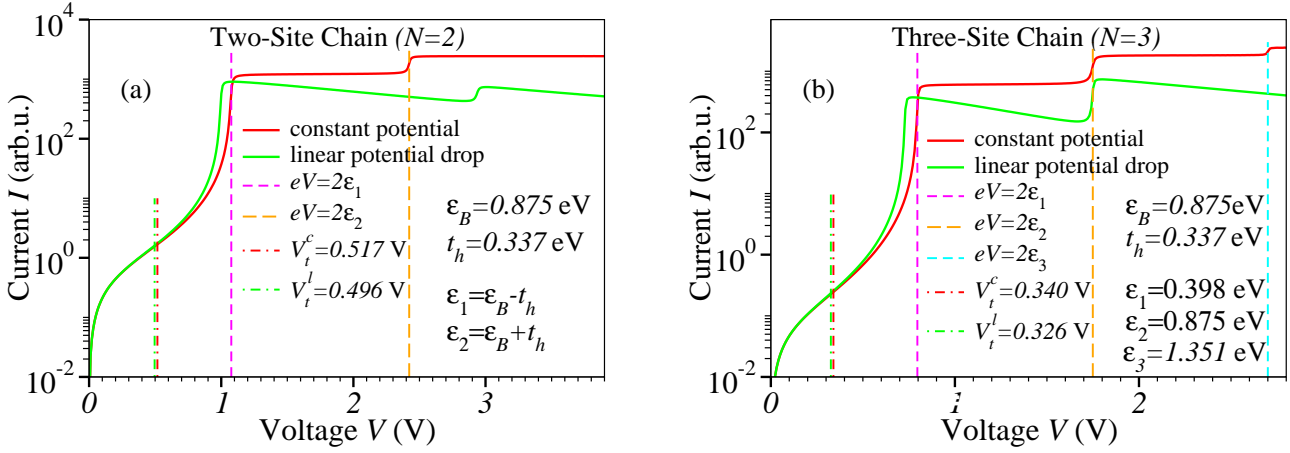

\centerline{\includegraphics[width=0.45\textwidth,angle=0]{fig_iv_n=2_long.eps}
\hspace*{5ex}
\includegraphics[width=0.45\textwidth,angle=0]{fig_iv_n=3_long.eps}
}
\caption{Tunneling current computed for molecular wires 
[model (iv), 
% eqn~(\ref{eq-ham-chains})] 
eqn~\ib{(S1)}]
for a constant (\figurename\ref{fig:tba-models}c) and linearly varying 
(\figurename\ref{fig:tba-models}d) potential. The linear potential drop has an impact 
at higher biases more pronounced than that visible in 
\figurename
\ib{S2}, % \ref{fig:n=2}, 
\ib{S3} % \figurename\ref{fig:n=3} 
and 
\ib{S4}; % \figurename\ref{fig:n=3}: 
displacements of the current step positions 
(which are located at on-resonance situations,
$eV = 2\varepsilon_j$, in the case of a constant potential)
and negative differential effects.
$\varepsilon_{j}$'s ($j\geq 1$) represent
the absolute values of the highest occupied orbital energies measured from the electrodes' Fermi energy. 
For a two-site chain, these values are $\varepsilon_{1,2} = \varepsilon_B \mp t_h$ (cf.~inset of panel (a)).
Notice the logarithmic scale on the y-axis.
}
\label{fig:ndr}
\end{figure*}
%%%%%%%%%%%%%%%%%%%%%%%%%%%%%%%%%%%%%%%%%%%%%%%%%%%%%%%%%%%%%%%%%%%%%%%%%%%%%%%%%%%%%%%%%%%%%%%%%% 

The impact of the spatial potential profile across a molecular junction is another issue worth to mention. 
Presumably, realistic potential profiles fall between the two situations depicted 
in panels (c) and (d) of \figurename\ref{fig:tba-models} 
(a constant potential and a linearly varying potential, respectively); so examining the differences
in the results obtained in these two cases may be taken to assess how important is to exactly know the 
actual potential profile in a real case. The results presented in 
\figurename
\ib{S2}, % \ref{fig:n=2}, 
\ib{S3} % \ref{fig:n=3}
and
\ib{S4} % \ref{fig:n=4} 
indicate 
a weak effect at biases not much larger than $V_t$. Obviously, this is a further plus of
transition voltage spectroscopy.
 
The manner in which the potential varies across a molecular junction is important only 
at biases substantially higher than $V_t$. This is shown in 
\figurename
\ib{S2}, % \ref{fig:n=2}, 
\ib{S3}, % \ref{fig:n=3}, 
\ib{S4} % \ref{fig:n=4}
and especially
in \figurename\ref{fig:ndr}, which, to the best of our knowledge, demonstrates a qualitatively new effect.
As visible there, instead of a current plateau occurring for a potential flat across the molecule
beyond bias values at which molecular orbital energies ($\varepsilon_{1,2,\ldots}$) 
become resonant to the Fermi level of one electrode ($e V_{res} = 2\varepsilon_{1,2,\ldots}$)
\cite{HaugJauho}, the current decreases for a linearly varying potential. 
We are not aware of a previous indication on such a negative differential resistance (NDR) effect
related to a specific potential variation across a molecular junction.     
Noteworthy, this is an NDR effect for uncorrelated transport computed 
within the limit of wide-band electrodes. Previous work 
on uncorrelated transport indicated finite electrode bandwidths and energy-dependent electrode 
density of states as possible sources of NDR.\cite{Baldea:2010e}
%%%%%%%%%%%%%%%%%%%%%%%%%%%%%%%%%%%%%%%%%%%%%%%%%%%%%%%%%%%%%%%%%%%%%%%%%%%%%%%%%%%%%%%%%%%%%%%%% 
\subsection{Gaussian transmission versus Lorentzian transmission}
%%%%%%%%%%%%%%%%%%%%%%%%%%%%%%%%%%%%%%%%%%%%%%%%%%%%%%%%%%%%%%%%%%%%%%%%%%%%%%%%%%%%%%%%%%%%%%%%% 
\label{sec:gaussian}
%%%%%%%%%%%%%%%%%%%%%%%%%%%%%%%%%%%%%%%%%%%%%%%%%%%%%%%%%%%%%%%%%%%%%%%%%%%%%%%%%%%%%%%%%%%%%%%%%
% The case of Gaussian transmission, 
% to which we refer in the end of 
% this section, is qualitatively different from models (i-v) examined above. 
% For models (i-v), the transition voltage
% $V_t$ increases with increasing energy offset $\varepsilon_B$. By contrast, 
% $V_t$ decreases with increasing energy offset $\varepsilon_B$ for Gaussian transmission 
% (\figurename\ref{fig:gaussian}a). 
% In fact, this is characteristic not only for Gaussian transmission, but also for general
% transmission given by eqn~(\ref{eq-T-delta}) with $\delta > 1$. 
%
The comparison between the cases of Lorentzian 
(obtained as a particular case of model (iv) for $N=1$, discussed in detail previously, e.g.,
ref.~\citenum{Baldea:2012a})
and Gaussian transmission (a particular case of eqn~(\ref{eq-T-delta}))
reveals that, even systems wherein the transport is dominated by a single energy level, 
may have qualitatively different physical properties.

\begin{figure*}[htb]
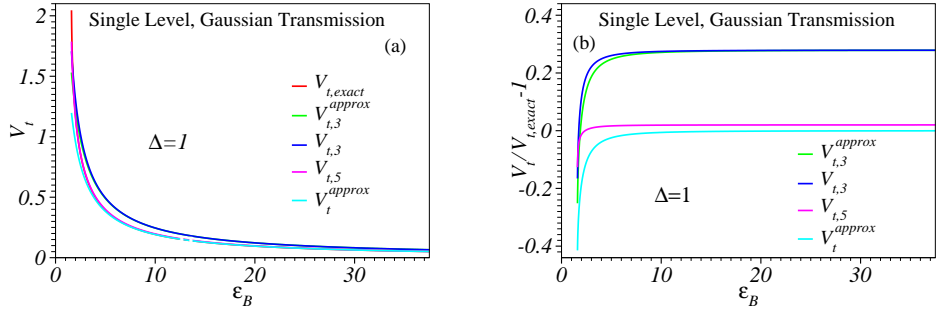

\centerline{\includegraphics[width=0.32\textwidth,angle=0]{fig_vt_gaussian_vs_e0.eps}
\hspace*{5ex}
\includegraphics[width=0.32\textwidth,angle=0]{fig_vt_relativ_gaussian_vs_e0.eps}}
% $ $\\[0ex]
\caption{Results obtained using the Gaussian transmission of 
% eqn~(\ref{eq-T-gauss}) 
eqn~\ib{(S4)}
showing the dependence on $\varepsilon_B$ of the transition voltage 
computed exactly ($V_t$), and using its approximate estimates: 
$V_{t,approx}$ of 
% eqn~(\ref{eq-vt-approx-gauss}), 
eqn~\ib{(S7)}
and $V_{t,3}^{approx}$, $V_{t,3}$ and $V_{t,5}$ obtained from eqn~(\ref{eq-vt3}), (\ref{eq-vt5}), 
\ib{(S8)} % (\ref{eq-c2-gauss}), 
and 
% (\ref{eq-c4-gauss}). 
\ib{(S9)}.
Notice that both $V_{t}^{approx}$ and $V_{t,5}$ represent good approximations for large values of the ratio $\varepsilon_B/\Delta$.}
\label{fig:gaussian-eB}
\end{figure*}

\begin{figure*}[htb]
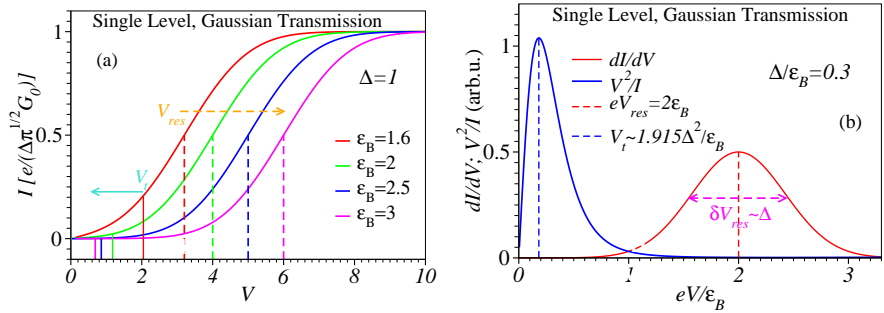

\centerline{\includegraphics[width=0.32\textwidth,angle=0]{fig_iv_gaussian.eps}
\hspace*{0ex}
\includegraphics[width=0.32\textwidth,angle=0]{fig_pvs_vResonance_gaussian.eps}}
% $ $\\[0ex]
\caption{Results obtained for a Gaussian transmission showing that, while the voltages where the $I-V$ curves (panel (a)) 
exhibit an inflection point (or maximum of the differential conductance $\partial I/\partial V$ of 
width proportional to $\Delta$, panel (b)) have a dependence similar to the case of a Lorentzian transmission (namely $e \vert V\vert  = 2\varepsilon_B$ corresponding to the alignment of the energy level $\varepsilon_B$ with the Fermi level of the biased electrodes), the transition voltage decreases with increasing $\varepsilon_B$, in contrast to the case of Lorentzian transmission, where it 
increases with  $\varepsilon_B$ ($e V_t = 1.154\,\varepsilon_B$)\cite{Baldea:2012a}.}
\label{fig:iv-fn-gaussian}
\end{figure*}

As visible in 
\figurename\ref{fig:gaussian-eB} and  
\ib{S5} % \ref{fig:gaussian-sigma} 
and 
% eqn~(\ref{eq-vt-approx-gauss}), 
eqn~\ib{(S7)}, 
$V_t$ strongly depends on the width parameter $\Delta$ and is inversely proportional to $\varepsilon_B$.
By contrast, in the case of Lorentzian transmission, 
$V_t$ linearly increases with $\varepsilon_B$ and is nearly independent of the width parameter $\Gamma$ 
as long as $\Gamma \ll \varepsilon_B$ [$e V_t = 2 \varepsilon_B/\sqrt{3} + \mathcal{O}\left(\Gamma/\varepsilon_B\right)^2$]. 
% \cite{Baldea:2010h,Baldea:2012a,Baldea:2012b,Baldea:2012g} 
\cite{Baldea:2010h,Baldea:2012a,Baldea:2012g} 
In fact, both the proportionality to  
$\varepsilon_B$ and the insensitivity to $\Gamma$-variations for $\Gamma \ll \varepsilon_B$ is 
a common feature
shared by the above models (iii-v), while a $V_t$ nearly proportional to $\varepsilon_B^{1/2}$ 
at large $\varepsilon_B$ for models (i) and (ii) is the consequence of the fact that the corresponding 
transmissions are similar to that of eqn~(\ref{eq-T-delta}) with $\delta=1/2$ 
(cf.~eqn~(\ref{eq-vt-ib}) and (\ref{eq-vt-approx-delta})).
Common features for these two types of transmissions are the inflection point of the $I-V$ curves
(``current steps'', \figurename\ref{fig:iv-fn-gaussian}a) located at biases $V_{res} = 2 \varepsilon_B / e$ where the 
level becomes resonant 
to electrodes' Fermi energy, and the fact that the associated maximum in the differential conductance has a width
proportional to the transmission widths; $\delta V_{res} \propto \Delta$ (\figurename\ref{fig:iv-fn-gaussian}b), similar to 
$\delta V_{res} \propto \Gamma$ for Lorentzian transmission.
The opposite dependence of $V_t$ on $\varepsilon_B$ for Lorentzian ($V_t \propto \varepsilon_B$) and Gaussian 
($V_t \propto 1/\varepsilon_B$) transmission is relevant for the correctly interpreting gating effects 
\cite{Wandlowski:08,Reed:09,Lo:15} or the impact of electrodes' work function.\cite{Frisbie:11}

Concerning the approximate estimates for $V_{t,exact}$, 
\figurename\ref{fig:gaussian-eB}b and 
\ib{S5}b % \ref{fig:gaussian-sigma}b 
indicate a behavior similar to that found in the above cases: $V_{t,5}$ represents an accurate estimate for the exact $V_{t,exact}$, while 
$V_{t,3}$ is quantitatively unsatisfactory. 

%%%%%%%%%%%%%%%%%%%%%%%%%%%%%%%%%%%%%%%%%%%%%%%%%%%%%%%%%%%%%%%%%%%%%%%%%%%%%%%%%%%%%%%%%%%%%%%%% % 
%%%%%%%%%%%%%%%%%%%%%%%%%%%%%%%%%%%%%%%%%%%%%%%%%%%%%%%%%%%%%%%%%%%%%%%%%%%%%%%%%%%%%%%%%%%%%%%%% % 
%%%%%%%%%%%%%%%%%%%%%%%%%%%%%%%%%%%%%%%%%%%%%%%%%%%%%%%%%%%%%%%%%%%%%%%%%%%%%%%%%%%%%%%%%%%%%%%%% % 
%%%%%%%%%%%%%%%%%%%%%%%%%%%%%%%%%%%%%%%%%%%%%%%%%%%%%%%%%%%%%%%%%%%%%%%%%%%%%%%%%%%%%%%%%%%%%%%%%%  
%%%%%%%%%%%%%%%%%%%%%%%%%%%%%%%%%%%%%%%%%%%%%%%%%%%%%%%%%%%%%%%%%%%%%%%%%%%%%%%%%%%%%%%%%%%%%%%%% 
\section{Asymmetric $\mathbf{I-V}$ curves described within the Newns-An\-der\-son mo\-del}
\label{sec:asym-NA}
%%%%%%%%%%%%%%%%%%%%%%%%%%%%%%%%%%%%%%%%%%%%%%%%%%%%%%%%%%%%%%%%%%%%%%%%%%%%%%%%%%%%%%%%%%%%%%%%% 
Let us briefly examine the case where, unlike those assumed throughout above, 
the $I-V$ curves are not symmetric under bias reversal, i.e., 
$I(-V)\neq -I(V)$. 
It is clear that the asymmetry with respect to bias polarity reversal cannot be satisfied within the
framework provided by eqn~(\ref{eq-i3}) and (\ref{eq-i5}); even powers in $V$ 
should also be added 
%%%%%%%%%%%%%%%%%%%%%%%%%%%%%%%%%%%%%%%%%%%%%%%%%%%%%%%%%%%%%%
\begin{eqnarray}
I(V) & = & I_{3} (V) + \mathcal{O}\left(V^4\right) = \nonumber \\
     & & G V \left(1 + c_{1} V + c_{2} V^2 \right) + \mathcal{O}\left(V^4\right) 
\label{eq-i3-even-odd} \\
I(V) & = & I_{5} (V) + \mathcal{O}\left(V^6\right) = G V \left(1 + c_{1} V + c_{2} V^2 \right . \nonumber \\
     & & + \left . c_{3} V^3 + c_{4} V^4 \right) + \mathcal{O}\left(V^6\right)
\label{eq-i5-even-odd} 
\end{eqnarray} 
%%%%%%%%%%%%%%%%%%%%%%%%%%%%%%%%%%%%%%%%%%%%%%%%%%%%%%%%%%%%%%%%%%%%%%%%%%%%%%%%%%%%%%%%%%%%%%%%% 
By including only the lowest even power (i.e.~$c_{1} \neq 0$, $c_{3} = 0$)\cite{Vilan:13}, it is possible 
to account for an asymmetry $I(V) \neq -I(-V)$. However, doing this, i.e.~using eqn~(\ref{eq-i3-even-odd}), 
has an important drawback.
By inserting eqn~(\ref{eq-i3-even-odd}) in eqn~(\ref{eq-vt}) one easily gets 
%%%%%%%%%%%%%%%%%%%%%%%%%%%%%%%%%%%%%%%%%%%%%%%%%%%%%%%%%%%%%%
\begin{equation}
\label{eq-equal-vt3+-}
V_{t,3+} = - V_{t,3-} = \frac{\varepsilon_B}{e \sqrt{c_{2}}}
\end{equation}
%%%%%%%%%%%%%%%%%%%%%%%%%%%%%%%%%%%%%%%%%%%%%%%%%%%%%%%%%%%%%%  
i.e., transition voltages of equal magnitudes for positive and negative bias polarities, 
and this holds no matter how large is $c_{1}$. Noteworthy, the asymmetry $I(V) \neq - I(-V)$ 
of an $I-V$ curve does not automatically imply $V_{t-} \neq - V_{t+}$. 
Experimental \cite{Beebe:06,Tan:10,Guo:11} and theoretical \cite{Baldea:2012a} results show 
that situations wherein $V_{t-} \neq - V_{t+}$ are physically relevant; this clearly 
demonstrates that the ``generic parabolic'' dependence of the 
conductance of eqn~(\ref{eq-i3-even-odd}) is an unsatisfactory approximation. 

To find the counterpart of eqn~(\ref{eq-vt5}) for the asymmetric case 
one has to solve an algebraic fourth order equation obtained by inserting eqn~(\ref{eq-i5-even-odd})
into eqn~(\ref{eq-vt}) 
%%%%%%%%%%%%%%%%%%%%%%%%%%%%%%%%%%%%%%%%%%%%%%%%%%%%%%%%%%%%%%
\begin{equation}
\label{eq-vt5-even-odd}
3 c_{4}^3 V_{t,5}^4 + 2 c_{3} V_{t,5}^3 + c_{2} V_{t,5}^2 = 1 
\end{equation}
%%%%%%%%%%%%%%%%%%%%%%%%%%%%%%%%%%%%%%%%%%%%%%%%%%%%%%%%%%%%%%
This yields asymmetric transition voltages $V_{t,5+} \neq -V_{t,5-}$.
Their analytical expressions are too long and will be omitted here.  

Starting from the models discussed above, asymmetric $I-V$ curves 
and $V_{t,+} \neq -V_{t,-}$ can be obtained by allowing a bias-induced 
energy shift 
%%%%%%%%%%%%%%%%%%%%%%%%%%%%%%%%%%%%%%%%%%%%%%%%%%%%%%%%%%%%%%
\begin{equation}
\label{eq-V-shift}
\varepsilon_B \to \varepsilon_B(V) = \varepsilon_B + \gamma e V
\end{equation}
%%%%%%%%%%%%%%%%%%%%%%%%%%%%%%%%%%%%%%%%%%%%%%%%%%%%%%%%%%%%%%
where $ -1/2 < \gamma < 1/2$ is a voltage division factor (see, e.g., 
ref.~\citenum{Baldea:2012a} and citations therein).     

\begin{figure*}[htb]
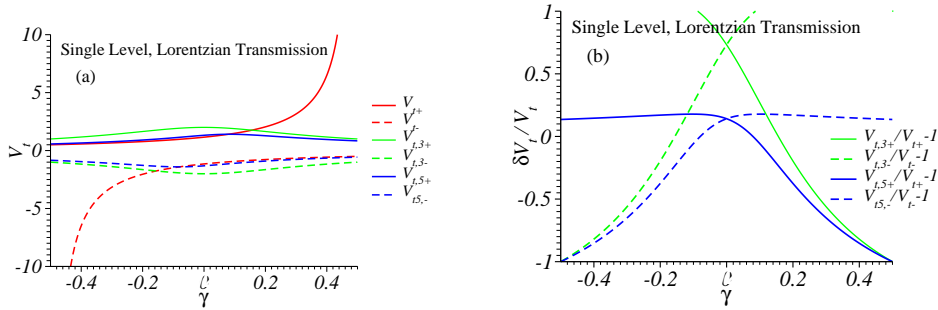

\centerline{\includegraphics[width=0.32\textwidth,angle=0]{fig_vt_vt3_vt5_lorentzian_vs_gamma.eps}
\hspace*{5ex}
\includegraphics[width=0.32\textwidth,angle=0]{fig_relative_vt_vt3_vt5_lorentzian_vs_gamma.eps}}
% $ $\\[0ex]
\caption{Results for a single level and Lorentzian transmission.
Transition voltages for positive and negative biases computed exactly $V_{t\pm}$ from 
eqn~\ib{(S11)}, % eqn~(\ref{eq-vt+-}),
and using the the third- or fifth-order expansions 
[$V_{t,3\mp}$, eqn~(\ref{eq-i3-even-odd}), and $V_{t,5\pm}$, eqn~(\ref{eq-i5-even-odd}), respectively].
Notice the incorrect prediction $V_{t,3+} = \vert V_{t,3-}\vert$.
The fifth-order expansion provides good estimates $V_{t,5\pm} \simeq V_{t\pm}$ only for small $\gamma$;
at larger $\gamma$, only one polarity is satisfactorily described (e.g., $V_{t,5-} \approx V_{t-}$ for $\gamma > 0$).}
\label{fig:vt-NA}
\end{figure*}

Although calculations are straightforward, 
being too lengthy, the counterpart of the formulas given for 
symmetric $I-V$ curves in the preceding section will not be given here. 
Instead, we will restrict ourselves to the Newns-Anderson (NA) model 
\cite{Newns:69b,MuscatNewns:78,Anderson:61,Schmickler:86,desjonqueres:96} 
within the wide-band limit, 
which assumes a single level of energy $\varepsilon_B(V)$ 
and Lorentzian transmission. For (off-resonant) situations of practical interest 
($\Gamma \ll \varepsilon_B$, biases up to $\sim 1.5\varepsilon_B/e$) 
% \cite{Baldea:2012b,Vuillaume:12a,Baldea:2012g,Vuillaume:12b,Tao:13,Fracasso:13,Vuillaume:15a,Lo:15,Vuillaume:15b}, 
\cite{Vuillaume:12a,Baldea:2012g,Vuillaume:12b,Tao:13,Fracasso:13,Vuillaume:15a,Lo:15,Vuillaume:15b}, 
exact formulae for the current and $V_{t\pm}$ have been deduced \cite{Baldea:2012a};
see eqn \ib{(S10)} % eqn~(\ref{eq-i-na-ib}) 
and \ib{(S11)} % (\ref{eq-vt+-}) 
in the {\si}.
%ESI exact formulas for the current and $V_{t\pm}$ have been deduced \cite{Baldea:2012a};
%%%%%%%%%%%%%%%%%%%%%%%%%%%%%%%%%%%%%%%%%%%%%%%%%%%%%%%%%%%%%%
%%%%%%%%%%%%%%%%%%%%%%%%%%%%%%%%%%%%%%%%%%%%%%%%%%%%%%%%%%%%%%
%ESI \begin{eqnarray}
%ESI \label{eq-i-na-ib} I & = & I (V; \gamma) = G V \frac{\varepsilon_B^2}{\left(\varepsilon_B + \gamma e V\right)^2 - (e V/2)^2} \nonumber \\
%ESI    & = & -I(-V; -\gamma) \neq - I(-V; \gamma) \nonumber \\
%ESI e V_{t+} & = & \frac{\varepsilon_B}{\sqrt{\gamma^2+3/4} - 2 \gamma} \\
%ESI e V_{t-} & = & - \frac{\varepsilon_B}{\sqrt{\gamma^2+3/4} + 2 \gamma} \label{eq-vt+-} 
%ESI \end{eqnarray}
%%%%%%%%%%%%%%%%%%%%%%%%%%%%%%%%%%%%%%%%%%%%%%%%%%%%%%%%%%%%%%
The fifth-order expansion in 
eqn \ib{(S10)} % eqn~(\ref{eq-i-na-ib}) 
straightforwardly yields the expansion coefficients entering eqn~(\ref{eq-i5-even-odd}) 
%%%%%%%%%%%%%%%%%%%%%%%%%%%%%%%%%%%%%%%%%%%%%%%%%%%%%%%%%%%%%%
\begin{eqnarray}
\tilde{c}_{1} & = & - 2 \gamma \ ; \ 
\tilde{c}_{2} = \frac{1}{4} + 3\gamma^2  
\label{eq-c1c2} \\
\tilde{c}_{3} & = & - \gamma \left( 1 + 4\gamma^2 \right) \ ; \ 
\tilde{c}_{4} = \frac{1}{16} + 5 \gamma^2 \left(\frac{1}{2} + \gamma^2 \right) 
\label{eq-c3c4} 
\end{eqnarray} 
%%%%%%%%%%%%%%%%%%%%%%%%%%%%%%%%%%%%%%%%%%%%%%%%%%%%%%%%%%%%%%%%%%%%%%%%%%%%%%%%%%%%%%%%%%%%%%%%% 
where $\tilde{c}_{j} \equiv c_{j} \left(\varepsilon_B/e\right)^j$ ($j=1\mbox{ to }4$).
By inserting the above expression in eqn~(\ref{eq-equal-vt3+-}) and 
(\ref{eq-vt5-even-odd}), 
the approximate values $V_{t,3\pm}$ and $V_{t,5\pm}$ for both bias polarities can be obtained.
The comparison with the exact $V_{t\pm}$ obtained from 
eqn~\ib{(S11)} % eqn~(\ref{eq-vt+-}) 
is depicted in 
\figurename \ib{S6} % \ref{fig:iv-NA} 
and \ref{fig:vt-NA}.
%%%%%%%%%%%%%%%%%%%%%%%%%%%%%%%%%%%%%%%%%%%%%%%%%%%%%%%%%%%%%%%%%%%%%%%%%%%%%%%%%%%%%%%%%%%%%%%%% 
\begin{figure*}[htb]
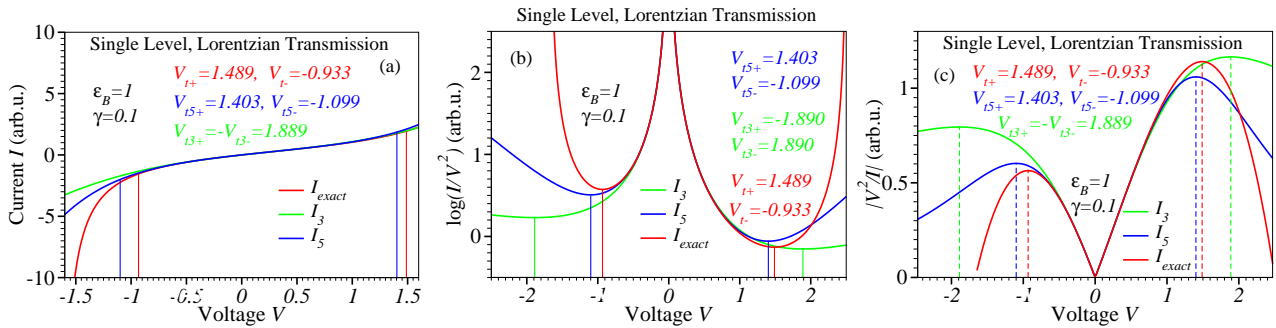

$ $\\[2ex]
\centerline{\includegraphics[width=0.31\textwidth,angle=0]{fig_iv_i3_i5_gamma_0.1_lorentzian.eps}\hspace*{0ex}
\includegraphics[width=0.31\textwidth,angle=0]{fig_fn_fn3_fn5_gamma_0.1_lorentzian.eps}\hspace*{0ex}
\includegraphics[width=0.31\textwidth,angle=0]{fig_pvs_i3_i5_gamma_0.1_lorentzian.eps}
}
% $ $\\[0ex]
\caption{Results for a single level and Lorentzian transmission. Whether computed exactly, 
using the third- or fifth-order expansions 
eqn \ib{(S10)}, % (eqn~(\ref{eq-i-na-ib}), 
(\ref{eq-i3-even-odd}), and (\ref{eq-i5-even-odd}), respectively) 
$I-V$ curves are asymmetric with respect to the origin 
for a nonvanishing voltage division factor $\gamma$ (panel (a)). The transition voltages
(minima of the Fowler-Nordheim quantity in panel (b)) computed exactly from 
eqn~\ib{(S11)} % eqn~(\ref{eq-vt+-})
for opposite polarities are of different magnitudes ($V_{t+} \neq \vert V_{t-}\vert $). 
The fifth-order expansion correctly describes this inequality
($V_{t,5+} \neq \vert V_{t,5-}\vert $), while the third-order
expansion incorrectly predicts $V_{t,3+} = \vert V_{t,3-}\vert $.
}
\label{fig:iv-NA}
\end{figure*}
%%%%%%%%%%%%%%%%%%%%%%%%%%%%%%%%%%%%%%%%%%%%%%%%%%%%%%%%%%%%%%%%%%%%%%%%%%%%%%%%%%%%%%%%%%%%%%%%%
%%%%%%%%%%%%%%%%%%%%%%%%%%%%%%%%%%%%%%%%%%%%%%%%%%%%%%%%%%%%%%%%%%%%%%%%%%%%%%%%%%%%%%%%%%%%%%%%% 
%%%%%%%%%%%%%%%%%%%%%%%%%%%%%%%%%%%%%%%%%%%%%%%%%%%%%%%%%%%%%%%%%%%%%%%%%%%%%%%%%%%%%%%%%%%%%%%%%

To end this subsection, lower order expansions (eqn~(\ref{eq-i3-even-odd}) and (\ref{eq-i5-even-odd}))
for the asymmetric $I-V$ curves considered above
appear to be more problematic than for symmetric cases. \figurename\ref{fig:vt-NA} shows 
that even the fifth-order expansion provides 
a satisfactory description for both bias polarities only for weak asymmetries ($\vert\gamma\vert \alt 0.1$). 
%%%%%%%%%%%%%%%%%%%%%%%%%%%%%%%%%%%%%%%%%%%%%%%%%%%%%%%%%%%%%%%%%%%%%%%%%%%%%%%%%%%%%%%%%%%%%%%%% 
\section{Case of molecular junctions based on oligophenylene dithiols}
%%%%%%%%%%%%%%%%%%%%%%%%%%%%%%%%%%%%%%%%%%%%%%%%%%%%%%%%%%%%%%%%%%%%%%%%%%%%%%%%%%%%%%%%%%%%%%%%% 
\label{sec:opds-parameters}
%%%%%%%%%%%%%%%%%%%%%%%%%%%%%%%%%%%%%%%%%%%%%%%%%%%%%%%%%%%%%%%%%%%%%%%%%%%%%%%%%%%%%%%%%%%%%%%%% 
Analytical expressions for current and conductance like those given above are 
very often utilized by experimentalists to fit their measurements. However, and this is one important point 
on which the present work wants to emphasize, these formulas are only valid under well 
defined conditions/restrictions, and utilization beyond these conditions makes no sense. To exemplify,
in this section we will analyze recent transport data obtained for molecular junctions 
based on oligophenylenes using silver electrodes Ag/OPDs/Ag containing up to four phenyl rings 
($1 \leq N \leq 4$).\cite{baldea:2015}

Experimental studies of OPD-based molecular junctions with silver electrodes containing $1\leq  N \leq 4$
phenyl rings yielded a value of the conductance attenuation factor
$\beta \simeq 1.56$ per ring;\cite{baldea:2015} for a phenyl ring size $d_1 = 4.3$\,{\AA}, 
this corresponds to $\tilde{\beta} \simeq 0.36\,\mbox{\AA}^{-1}$ \cite{baldea:2015}. 
This $\beta$-estimate can be invoked to immediately rule out the tunneling barrier picture underlying 
models (i) and (ii) discussed above as valid framework to analyze the charge transport in these junctions. 
Out of the phenylene dithiol species studied in ref.~\citenum{baldea:2015},
it is only the four-member species, $d \to d_4 = 4 d_1$, that satisfies condition 
of eqn~(\ref{eq-validity-simmons}), $\tilde{\beta} d = 6.24 > 4$. In fact, as discussed in detail recently 
% \cite{Baldea:2012c,Baldea:2012e,Baldea:2012f}, 
\cite{Baldea:2012c,Baldea:2012e}, 
the Simmons model in its original 
formulation \cite{Simmons:63} does not account for the fact that, in molecular junctions, 
electron motion is laterally confined. 
% One should mention that even in CP-AFM junctions, 
% which are not characterized by atomically sharp tips, the lateral constriction is relevant; 
% electron motion is highly anisotropic, basically proceeding along the molecules 
% \cite{Baldea:2012c,Baldea:2012e,Baldea:2012f}. 
% Furthermore,
% the fact that interactions between molecules do not notably affect the transport in the CP-AFM junctions 
% based on oligophenylenes is supported by the experimental transport data \cite{baldea:2015}. 
When lateral constriction is incorporated into theory, the condition 
becomes more restrictive. Instead of eqn~(\ref{eq-validity-simmons}), 
one should apply eqn~(\ref{eq-validity-ib}), which is invalidated even for $N=4$.

% The contact conductance $G_{c}$ is another critical issue within the tunneling barrier description even at qualitative level.  With respect to the transmission computed by solving the Schr\"odinger equation, the transmission obtained within WKB-based approximations like that underlying the Simmons model misses a prefactor, whose importance has been emphasized recently,\cite{Forbes:08,Baldea:2012f} which is accounted for in eqn~(\ref{eq-G-ib}).\cite{Baldea:2012c} Because within errors there is no significant difference in the $\beta$-values for OPDs junctions with Ag, Au, and Pt electrodes (work functions $\Phi_{Ag} = 4.25$\,eV, $\Phi_{Au} = 5.2$\,eV, and $\Phi_{Pt} = 5.65$\,eV), eqn~(\ref{eq-G-ib}) predicts $G_{c}^{Au}/G_{c}^{Ag} = 0.82$ and $G_{c}^{Pt}/G_{c}^{Au} = 0.92$.  The values are at odds with the experimental ones $G_{c}^{Au}/G_{c}^{Ag} \simeq 13$ and $G_{c}^{Pt}/G_{c}^{Az} \simeq 6$.\cite{baldea:2015} 

The presently employed parameter $\varepsilon_B$ represents an effective barrier height,
which also embodies image effects. 
According to eqn~(\ref{eq-alpha}) and (\ref{eq-beta-tilde}), 
to ``explain'' a certain $\beta$-value adjusted to fit experimental data, both $\varepsilon_B$ and the
the effective mass $m^\ast$ can empirically be ``adjusted'' to make theory to ``agree'' with experiment.
However, what cannot be ``manipulated'' in this way is the $\beta$-value itself. 
The restriction expressed by eqn~(\ref{eq-validity-ib}) represents the mathematical condition for  
a valid description within the tunneling barrier model, and it cannot be modified
whatever ``appropriate'' is the choice of $\varepsilon_B$ and $m^\ast$. 
% Likewise, since the $\tilde{\beta}$-value is fixed by experiments, eqn~(\ref{eq-G-ib}) does not allow any ``manipulation'' of the dependence of $G_{c}$ on the work function $\Phi$. 

The argument based on the (too small) $\beta$-value also demonstrates that a description based on 
the superexchange limit of tunneling,
which underlines model (iii),\cite{Mujica:01} is impossible. 
% For $\beta = 1.58$, eqn~(\ref{eq-G-mr}) yields $t_h/\varepsilon_B = 0.454$, 
For $\beta = 1.56$, eqn~(\ref{eq-G-mr}) yields $t_h/\varepsilon_B = 0.458$,
which is not much smaller than unity, as required by eqn~(\ref{eq-validity-mr}). 
Again, whatever ``appropriate'' the adjustment of the parameters $\varepsilon_B$ and $t_h$,
they cannot be chosen to satisfy the condition required by theory for a valid 
superexchange mechanism, because the experimental data ``fix'' the value of $\beta$ and 
thence via eqn~(\ref{eq-G-mr}) the ratio $t_h/\varepsilon_B$.
So, one should go beyond the superexchange limit. 

The most straightforward 
generalization is to use model (iv) for a chain with  
having one level per monomer, for which theory does not impose restrictions to the  
ratio between $\varepsilon_B$ and $t_h$. 
Within a microscopic description based on model (iv),
$t_h$ represents the hopping integral coupling the two HOMOs of two adjacent benzene rings. 
It can be deduced from the difference 
between the ionization energies $I_1$ and $I_2$ of benzene and biphenyl: $2 t_h = I_1 - I_2$.
%
% c6h6:     I1: CCSD: 9.21904 OVFG -9.197 D-DFT 9.23491
% biphenyl: I2: CCSD: 8.217 OVGF -8.204 D-DFT 8.04327
%
Quantum chemical calculations based on refined ab initio methods, EOM-CCSD 
(equation-of-motion coupled clusters singles and doubles), as implemented 
in CFOUR \cite{cfour} and OVGF (outer valence Green's functions) \cite{Cederbaum:75,Cederbaum:77,Schirmer:84}
% and $\Delta$-DFT \cite{Baldea:2013b,Baldea:2014c}, 
done with GAUSSIAN 09,\cite{g09} allowed us 
to reliably determine $I_1$ and $I_2$. The values are 
% $I_1 = 9.219; 9.197; 9.235$\,eV 
$I_1 = 9.219$\,eV and $9.197$\,eV, and 
% $I_2 = 8.217; 8.204; 8.043$\,eV, 
$I_2 = 8.217$\,eV and $8.204$\,eV, respectively.
They allow to compute the hopping integral needed for calculations based on model (iv):
$t_h = 0.501$\,eV and $0.497$\,eV, respectively. 
The agreement between these two $t_h$-values demonstrates the reliability
of the ab initio estimates. 
The direct ab initio determination of $\varepsilon_B$ 
(relative alignment between HOMO and electrodes' Fermi level) is an issue too difficult to be addressed here. 
It can be determined from the experimental value
$\left . V_t \right\vert_{N=1}\simeq 1.15$\,V for Ag/OPD1/Ag.\cite{Frisbie:11,baldea:2015} 
The equation $\varepsilon_B = e V_t \sqrt{3}/2\ $ \cite{Baldea:2012a}
yields $\varepsilon_B = 0.996$\,eV.
Exact currents for the Hamiltonian of 
eqn~\ib{(S1)} % eqn~(\ref{eq-ham-chains}) 
can be computed numerically,
and this allows to obtain $V_t$ for various $N$'s. 
Results based on model (iv) using this ab initio estimated $t_h=0.5$\,eV and adjusting $\varepsilon_B$
to the value $\varepsilon_B = 0.996$\,eV fixed by the experimental value $\left . V_t \right\vert_{N=1}\simeq 1.15$\,V \cite{baldea:2015} 
are completely unacceptable; not even 
an exponential decay of the conductance with increasing size $N$ can be obtained.
% (\figurename\ref{fig:chains-vt1-opds}a).

\begin{figure*}[htb]
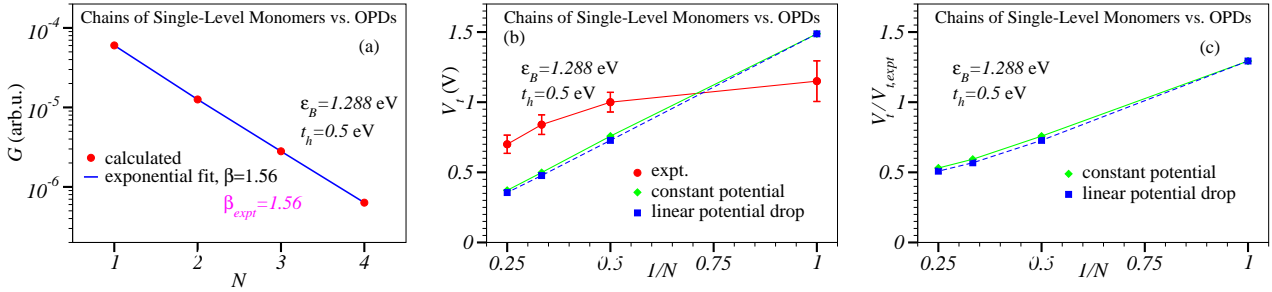

% 
% \centerline{\includegraphics[width=0.3\textwidth,angle=0]{fig_beta_1.58_chains_eS_1.2975_ab_initio_tS_0.5_opds.eps}
\centerline{\includegraphics[width=0.3\textwidth,angle=0]{fig_beta_1.56_Ag_chains_eS_1.2975_ab_initio_tS_0.5_opds.eps}
\hspace*{1ex}
\includegraphics[width=0.3\textwidth,angle=0]{fig_vt_beta_1.56_Ag_chains_eS_1.288_ab_initio_tS_0.5_opds.eps}
\hspace*{1ex}
\includegraphics[width=0.3\textwidth,angle=0]{fig_vt_relativ_beta_1.56_Ag_chains_eS_1.288_ab_initio_tS_0.5_opds.eps}}
% $ $\\[0ex]
\caption{Results for model (iv) using the ab initio value $t_h=0.5$\,eV and 
the value $\varepsilon_B = 1.288$\,eV obtained by fitting 
the experimental conductance tunneling attenuation coefficient $\beta=1.56$ (panel (a))\cite{baldea:2015}  
The agreement between the calculated $V_t$-values (panel (b)) and experiment is poor.
The spatial potential profile across these junctions
do not notably affect $V_t$, as illustrated by the results obtained for constant and linearly varying potentials 
(panel (c)).
The points (panels (b) and (c)) and error bars (panel (b)) represent experimental results for 
CP-AFM Ag/OPDs/Ag junctions.\cite{baldea:2015}
} 
\label{fig:chains-opds-fixed-t_h}
\end{figure*}

In view of this disagreement, we have attempted to keep only one of the above parameters 
(either $t_h=0.5$\,eV or $\varepsilon_B = 0.996$\,eV) and to determine the other by 
fitting the experimental value $\beta=1.56$.\cite{baldea:2015}
Fixing $\varepsilon_B = 0.996$\,eV 
yields $t_h=0.387$\,eV, while fixing $t_h=0.5$\,eV requires a value $\varepsilon_B = 1.288$\,eV.
The results obtained in this way are depicted in 
\figurename
\ib{S6} % \ref{fig:chains-opds-fixed-epsilon_B} 
and 
\ref{fig:chains-opds-fixed-t_h}, respectively,
and they show that none of these empirical changes can make
model (iv) to provide a satisfactory description. 
Differences between the cases of a flat potential
and a linearly varying potential are insignificant 
(\figurename
\ib{S6}c % \ref{fig:chains-opds-fixed-epsilon_B}c 
and \ref{fig:chains-opds-fixed-t_h}c); so, the lack of information
of the real spatial potential profile across junctions cannot be advocated as possible source of this disagreement. 
%%%%%%%%%%%%%%%%%%%%%%%%%%%%%%%%%%%%%%%%%%%%%%%%%%%%%%%%%%%%%%%%%%%%%%%%%%%%%%%%%%%%%%%%%%%%%%%%% 
%%%%%%%%%%%%%%%%%%%%%%%%%%%%%%%%%%%%%%%%%%%%%%%%%%%%%%%%%%%%%%%%%%%%%%%%%%%%%%%%%%%%%%%%%%%%%%%%%%

A description based on model (v), which considers the phenyl rings
explicitly, represents the highest reasonable refinement of a tight-binding approach to 
charge transport in OPDs junctions.
To determine, $t_i$, which within this model is a characteristic of a ring, we use the
lowest singlet excitation energy $E_{exc}$; it can be expressed as $E_{exc} = 2 t_i$. 
Using aug-cc-pVDZ basis sets, we have determined $E_{exc}$ from  EOM-CCSD \cite{cfour}
and SAC-CI (symmetry adapted cluster/configuration interaction) 
\cite{g09} calculations; the values are $E_{exc}=5.154$\,eV and $E_{exc}=5.051$\,eV, respectively.    
They are in good agreement with the experimental values: $4.9$\,eV 
\cite{Callomon:66,Lassettre:68} and $5.0$\,eV.\cite{Doering:69}
So, we use the value $t_i = 2.5$\,eV, which also agrees with the overall description of the 
excitation spectrum of benzene 
% \cite{Baldea:2007} 
as well as with polyacetylene data 
% \cite{Baldea:2014b}. 
\cite{Baldea:2007}.
To determine $t$, we use again the difference between the lowest ionization energies 
of benzene and biphenyl. 
(Notice that the parameter $t$ of model (v), which is the hopping integral between 
neighboring C sites belonging to adjacent phenyl rings (\figurename\ref{fig:tba-models}b), 
is different from the parameter $t_h$ of model (iv), 
which is the hopping integral between HOMOs of adjacent phenyl rings (\figurename\ref{fig:tba-models}a).) 
To reproduce $I_1 - I_2 \simeq 1$\,eV (see above), a value $t = 3.677$\,eV is needed.  
The very fact that $t=3.677$\,eV is larger than $t_i=2.5$\,eV is unphysical; the hopping 
integral $t$ associated with a single C-C bond cannot be larger than the hopping integral $t_i$ 
associated to neighboring carbon atoms in an aromatic ring characterized by C-C distances 
shorter than of a single C-C bond (\figurename\ref{fig:tba-models}b).
It is a clear expression that, parameters of tight-binding models (however refined they are) 
cannot be adjusted to satisfactorily oligophenylene molecules. This finding is in line with 
recent studies demonstrating limitations of tight-binding approaches for molecules of interest 
for molecular electronics.\cite{Baldea:2014c}

\begin{figure*}[htb]
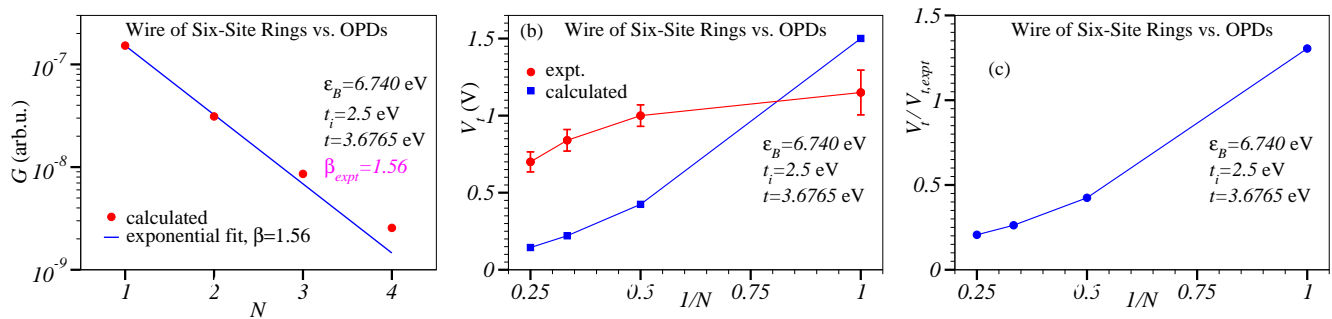

\centerline{
\includegraphics[width=0.32\textwidth,angle=0]{fig_beta_1.56_Ag_eS_6.74_tS_2.5_tB_3.6765_rings_opds.eps} \hspace*{0ex}
\includegraphics[width=0.32\textwidth,angle=0]{fig_vt_beta_1.56_Ag_eS_6.74_tS_2.5_tB_3.6765_rings_opds.eps} \hspace*{0ex}
\includegraphics[width=0.32\textwidth,angle=0]{fig_vt_relativ_beta_1.56_Ag_eS_6.74_tS_2.5_tB_3.6765_rings_opds.eps}
}
% $ $\\[0ex]
\caption{Results for the conductance (panel (a)) and transition voltage (panel (b)) computed 
within model (v) 
using the values $t_i=2.5$\,eV and $t=3.677$\,eV deduced ab initio and 
adjusting $\varepsilon_B=6.74$ to fit the tunneling attenuation factor $\beta=1.56$.
In panel (b), the points and error bars represent experimental 
results for CP-AFM Ag/OPDs/Ag junctions.\cite{baldea:2015}}
\label{fig:ab-initio-rings-opds}
\end{figure*}

Using the ab initio values $t=3.677$\,eV and $t_i=2.5$\,eV and adjusting $\varepsilon_B$
to fit $\left . V_t \right\vert_{N=1} \simeq 1.15$\,V yields $\varepsilon_B = 3.433$\,eV. This parameter 
set is not even able to qualitatively describe the exponential decay with $N$ of the conductance.
Therefore, we have also considered two alternative parameter sets:
$\varepsilon_B = 6.74$\,eV, $t_i=2.5$\,eV, and $t=3.677$\,eV 
(\figurename\ref{fig:ab-initio-rings-opds})
and
$\varepsilon_B = 3.433$\,eV, $t_i=2.5$\,eV, and $t=0.964$\,eV 
(\figurename \ib{S7}). % \ref{fig:rings-opds}).
To get the first set, instead of fitting $\left . V_t \right\vert_{N=1}\simeq 1.15$\,V (as done above), 
we have adjusted $\varepsilon_B$ by imposing $\beta=1.56$. To obtain the second set,
we have adjusted $\varepsilon_B$ and $t$ to fit the experimental values $\beta=1.56$ and 
$\left . V_t \right\vert_{N=1}\simeq 1.15$\,V. Notice that the value $t=0.964$\,eV is much smaller 
not only than the ab initio estimate $t=3.677$\,eV and also smaller than $t_i = 2.5$\,eV, which cannot 
be explained by the fact that a C-C single bond is slightly shorter than an aromatic one.\cite{ssh:2}
As visible in \figurename\ref{fig:ab-initio-rings-opds} and 
\ib{S7}, % \ref{fig:rings-opds},
none of these two sets 
provides a satisfactory description.
%%%%%%%%%%%%%%%%%%%%%%%%%%%%%%%%%%%%%%%%%%%%%%%%%%%%%%%%%%%%%%%%%%%%%%%%%%%%%%%%%%%%%%%%%%%%%%%%%% 
%%%%%%%%%%%%%%%%%%%%%%%%%%%%%%%%%%%%%%%%%%%%%%%%%%%%%%%%%%%%%%%%%%%%%%%%%%%%%%%%%%%%%%%%%%%%%%%%%%  

The (in)accuracy of $V_t$-estimates based on the third- and fifth-order expansions of 
eqn~(\ref{eq-i3}) and (\ref{eq-i5}) using parameter values specific for OPDs junctions, 
which is illustrated in 
\figurename \ib{S8}, % \ref{fig:vt3-vt5-chains-rings-opds}, 
is comparable to that 
encountered in the previous situations 
(\figurename\ref{fig:simmons}, 
\ib{S1}, % \ref{fig:sinh}, 
\ref{fig:mr}, 
\ref{fig:gaussian-eB} 
and 
\ib{S5}). % \ref{fig:gaussian-sigma}).

To conclude this part, neither representing OPDs junctions 
as chains of phenyl rings merely considering coupled HOMO's of neighboring rings within model (iv) 
nor a full treatment of the coupled rings at tight binding level is able to provide
an acceptable description of the transport experiments.\cite{baldea:2015}
Limitations of tight binding descriptions of molecules of interest for molecular electronics have 
been recently pointed out.\cite{Baldea:2014c}

The data collected in \ref{table:gaussian-Ag} demonstrate why a description based on a single level and 
Gaussian transmission should also be ruled out 
for OPDs junctions. The difference between the values $\varepsilon_B$ deduced from 
transport data \cite{baldea:2015} and from ultraviolet photoelectron spectroscopy (UPS) \cite{Frisbie:11}
for Ag/OPD1/Ag is enormous in the only case ($N=1$) where the latter is available. In addition, 
$\varepsilon_B$, which represents the HOMO energy offset relative to the Fermi level, 
is found to increase with increasing $N$, which is completely unphysical.\cite{baldea:2015}
%%%%%%%%%%%%%%%%%%%%%%%%%%%%%%%%%%%%%%%%%%%%%%%%%%%%%%%%%%%%%%%%%%%%%%%%%%%%%%%%%%%%%%%%%%%%%%%%% 
\begin{table} % [h!]
{\footnotesize % \small
\begin{center}
\begin{tabular*}{0.4\textwidth}{@{\extracolsep{\fill}}ccccc}
$N$ & $V_t$ $^{\mbox{a}}$ & $\Delta$ & $\varepsilon_B$ & $\varepsilon_B^{expt}$\\
\hline
1 &   1.15 & 1.273 & 3.061 & 1.1 $^{\mbox{b}}$ \\
2 &   1.00 & 1.265 & 3.378 &     \\
3 &   0.84 & 1.212 & 3.615 &     \\
4 &   0.70 & 1.118 & 3.639 &     \\
\hline
\end{tabular*}
\caption{Results obtained by modeling Ag/OPDs/Ag by assuming a Gaussian transmission, 
eqn~\ib{(S4)}. % eqn~(\ref{eq-T-gauss}).
$^{\mbox{a}}$: experimental values \cite{baldea:2015};  
$^{\mbox{b}}$: result of ultraviolet photoelectron spectroscopy (UPS) \cite{Frisbie:11}.
}
\label{table:gaussian-Ag}
\end{center}
}
\end{table}
%%%%%%%%%%%%%%%%%%%%%%%%%%%%%%%%%%%%%%%%%%%%%%%%%%%%%%%%%%%%%%%%%%%%%%%%%%%%%%%%%%%%%%%%%%%%%%%%%
%%%%%%%%%%%%%%%%%%%%%%%%%%%%%%%%%%%%%%%%%%%%%%%%%%%%%%%%%%%%%%%%%%%%%%%%%%%%%%%%%%%%%%%%%%%%%%%%%
\ib{
\section{Additional remarks}
\label{sec:remarks}
%%%%%%%%%%%%%%%%%%%%%%%%%%%%%%%%%%%%%%%%%%%%%%%%%%%%%%%%%%%%%%%%%%%%%%%%%%%%%%%%%%%%%%%%%%%%%%%%%
To avoid possible misunderstandings related to the present paper, 
a few remarks are in order, however.

The above considerations refer to \emph{existing} models, which are 
currently used for interpret experimental data 
for the charge transport \emph{via} tunneling in molecular junctions.
Our main aim was to present a list of formulae that can be used for data processing
along with the pertaining applicability conditions. 
This paper is intended as a working instrument enabling to check whether
certain experimental transport data are compatible or not with one of the 
existing models. By presenting the benchmark case of 
OPD-based junctions, we mainly aimed at illustrating how experimental data could/should
be utilized to (in)validate a certain theoretical model, not merely checking whether 
$I-V$ measured curves can be fitted without checking whether the values of the 
fitting parameters are acceptable. An exhaustive analysis of transport data available in 
the literature for various molecular junctions within the presently considered models
is beyond the present aim; this may make the object of a (review) paper of interest on its own. 

The presently considered models disregard important physical effects (\emph{e.g.}, 
details of interface 
molecule-electrode couplings or metal-induced gap states). It may be entirely possible that 
none of these models is entirely satisfactory just because such effects are significant.
A corresponding extension of the theoretical models to provide experimentalists 
with simple formulae enabling data processing/interpretation, obviating demanding microscopic   
transport calculations remains a desirable task for the future.

In this paper we did not discuss the (off-resonant limit ($\Gamma \ll \varepsilon_B$) 
of the) Newns-Anderson model in much detail.
Relevant formulae 
(eqn~\ib{(S10)} and \ib{(S11)}) % (eqn (\ref{eq-i-na-ib}) and (\ref{eq-vt+-})) 
are given in the {\si}.
As shown in recent studies \cite{Baldea:2013b,Baldea:2014e,Baldea:2015b,baldea:2015}, 
cases where this description applies and is justified microscopically exist. 
There, the transition voltage can be used
to directly extract the orbital energy offset using a formula 
($ e V_t = 2 \varepsilon_B / \sqrt{3} = 1.155\, e V_t$ for $\gamma = 0$ \cite{Baldea:2012a})
close to that initially claimed ($e V_t = \varepsilon_B $)\cite{Beebe:06}.
However, the results presented for the various models analyzed above
have demonstrated that
$V_t$ cannot be uncritically used to straightforwardly deduce the energy alignment 
of the dominant molecular orbital. This does not diminish the importance of 
$V_t$. As emphasized in Section \ref{sec:Vt}, 
$V_t$ is an important property characterizing the nonlinear transport. 
By studying its behavior across homologous molecular classes (\emph{i.e.}, 
varying the molecular size $d$ or $n$) or under gating 
(\emph{i.e.}, varying $\varepsilon_B$), the (in)applicability 
of a certain model can be concluded.
Moreover, $V_t$ turned out to be a key quantity, as it allowed to reveal that charge transport
across different experimental platforms and different molecular species exhibit a
universal behavior, which can be even formulated as a law of corresponding states (LCS) 
free of any empirical parameters \cite{Baldea:2015b}.

In order to demonstrate that the cubic polynomial approximation is insufficiently accurate 
to describe $V_t$, we have presented in 
\figurename\ref{fig:generic-parabola}a a raw $I-V$ trace measured on a CP-AFM junction 
\cite{baldea:2015}. Making the point in connection with 
\figurename\ref{fig:generic-parabola} is only possible by using neat, smooth curves;
typical raw $I-V$ curves measured for single-molecule junctions are blurry and therefore
inadequate for this purpose.
A comparison between junctions consisting of a single molecule and 
a bundle of molecules is of interest on its own; still, we note that, as long as the 
charge transfer occurs \emph{via} off-resonant tunneling, the similarity of the 
transport properties of these two types of junctions is expressed
by the aforementioned LCS \cite{Baldea:2015b}. Considering transport through 
CP-AFM junctions within the models discussed above amounts to neglect proximity 
effects due to other (identical) molecules in the bundle. This assumption may certainly fail
in some cases. However, the fact that 
OPD-based CP-AFM junctions obey this LCS  \cite{Baldea:2015b} can be taken   
as a strong indication that the failure of the various models in case of these molecular devices
discussed in Section \ref{sec:opds-parameters} 
is not related to the approximate description in terms of of a bundle of independent molecules. 
}
%%%%%%%%%%%%%%%%%%%%%%%%%%%%%%%%%%%%%%%%%%%%%%%%%%%%%%%%%%%%%%%%%%%%%%%%%%%%%%%%%%%%%%%%%%%%%%%%% 
\section{Conclusion}
\label{sec:conclusion}
%%%%%%%%%%%%%%%%%%%%%%%%%%%%%%%%%%%%%%%%%%%%%%%%%%%%%%%%%%%%%%%%%%%%%%%%%%%%%%%%%%%%%%%%%%%%%%%%% 
With the manifest aim of providing experimental colleagues a comprehensive working framework 
enabling them to process and interpret measurements of transport by tunneling in molecular junctions, 
this paper have presented a detailed collection of analytical formulae, emphasizing on the fact 
(less discussed in the literature) 
that these formulae only hold if specific conditions of applicability
are satisfied, which often impose severe restrictions on the model parameters. 
From a more general perspective, the theoretical results reported above have demonstrated that:

(i)
The often accepted idea of a generic parabolic $V$-dependence of the conductance 
as fingerprint of transport via tunneling emerged from studies based on 
high and wide energy barriers.
This picture misses a microscopic foundation for tunneling across  
molecules characterized by discrete energy levels. 
At biases of experimental interest, within all the models examined here, 
the description based on a third-order expansion $I=I(V)$
of transport measurements in molecular junctions 
is insufficient for quantitative purposes. To illustrate, we have shown this by systematically 
analyzing the fifth-order expansions, which turned out to be reasonably accurate at least for 
biases up to the transition voltages.

(ii)
Merely adjusting model parameters (e.g., within a tunneling barrier picture by claiming 
renormalization effects due to image charges or effective mass) does not suffice to 
``made'' a model valid for describing a specific molecular electronic device. On one side, 
models are normally valid only under specific conditions, which
these parameters should satisfy.
On the other side, the parameter values deduced from fitting experimental data should be consistent
with ab initio estimates. Parameters for tight-binding models can be easily estimated via reliable 
ab initio quantum chemical calculations, as illustrated in this study.

(iii)
The experimentally measured values of the $\beta$-tunneling coefficient of the low bias resistance
can be inferred for quickly assessing the inapplicability of certain tunneling models.
This turned out to be the case for OPDs junctions,
whose (too small) value ($\beta=1.56$ or $\tilde{\beta}=0.36\,\mbox{\AA}^{-1}$)
deduced from experiments is incompatible with descriptions based on 
tunneling barrier or superexchange mechanism. Because molecular junctions based 
on other aromatic aromatic species have often even smaller $\beta$'s 
($\beta \approx 0.2\,\mbox{\AA}^{-1}\ $ \cite{Frisbie:11,Reddy:12a}), 
descriptions based on those models should be excluded. 
In such cases, uncritical application of the mathematical formula of $I$ vs.~$V$ is meaningless, even 
if the measured $I-V$ curves can be satisfactorily fitted.

(iv)
For biases not too much higher than the transition voltage, 
a realistic description of the spatial potential profile across a molecular junction appears considerably
less important that usually claimed. This is no longer the case at high biases, where the spatial potential
profile may be responsible for qualitatively new phenomena, e.g., 
negative differential resistance (cf.~\figurename\ref{fig:ndr}). 

(v)
The fact that the estimate $V_{t,5} \approx V_{t}$ based on the 
fifth-order $I(V)$-expansion turned out to be acceptable in many cases
can be of practical help; eqn~(\ref{eq-i5}) (or eqn~(\ref{eq-i5-even-odd}))
can be employed to process noisy experimental 
$I-V$ curves, for which straightforwardly redrawing
measurements as $\log(I/V^2)$ vs.~$V$ (or $V^2/I$ vs.~$V$) 
may yield substantial uncertainties to determine the minimum (or maximum) 
position. Fitting $I-V$ curves with fifth-order polynomials can be used 
to extract the coefficients $c_{2,3,4}$ and to estimate $V_t \approx V_{t,5}$
by means of eqn~(\ref{eq-vt5}) or (\ref{eq-vt5-even-odd}).

(vi) Finally, we refer to situations where a successful description based on a 
single level \emph{and} 
Lorentzian transmission (also known as the Newns-Anderson model) has been concluded 
\cite{Baldea:2013b,Baldea:2014e,baldea:2015}. The fact that the $I-V$ curves could be very 
well fitted within the Newns-Anderson framework was not the only argument 
leading to that conclusion; equally important was that the energy offset $\varepsilon_B$ deducing
from fitting the $I-V$ data has been correlated with ab initio estimates \cite{Baldea:2014e,baldea:2015} 
and/or independent experimental information.\cite{Baldea:2013b} Even if the transport would have been
dominated by a single level, the conclusion of those studies would have not emerged in case of, e.g.~a Gaussian transmission, because of the completely different dependence of $V_t$ on $\varepsilon_B$.  

With the (few) examples mentioned in the preceding paragraph, we want to end by reiterating
an idea already presented in Introduction: while attempting to make the community more aware of
the limits of applicability of the various models utilized, 
the present paper did by no means intend to challenge the overall usefulness 
of model-based studies in gaining conceptual insight into the charge transport by tunneling at nanoscale.
%%%%%%%%%%%%%%%%%%%%%%%%%%%%%%%%%%%%%%%%%%%%%%%%%%%%%%%%%%%%%%%%%%%%%%%%%%%%%%%%%%%%%%%%%%%%%%%%% 
%%%%%%%%%%%%%%%%%%%%%%%%%%%%%%%%%%%%%%%%%%%%%%%%%%%%%%%%%%%%%%%%%%%%%%%%%%%%%%%%%%%%%%%%%%%%%%%%% 
\section*{Acknowledgement}
Financial support provided by the
Deu\-tsche For\-schungs\-ge\-mein\-schaft
(grant BA 1799/2-1) is gratefully acknowledged.
%%%%%%%%%%%%%%%%%%%%%%%%%%%%%%%%%%%%%%%%%%%%%%%%%%%%%%%%%%%%%%%%%%%%%%%%%%%%%%%%%%%%%%%%%%%%%%%%
%
\balance
%%%%%%%%%%%%%%%%%%%%%%%%%%%%%%%%%%%%%%%%%%%%%%%%%%%%%%%%%%%%%%%%%%%%%%%%%%%%%%%%
%%%%%%%%%%%%%%%%%%%%%%%%%%%%%%%%%%%%%%%%%%%%%%%%%%%%%%%%%%%%%%%%%%%%%%%%%%%%%%%%
\renewcommand\refname{Notes and references}
\providecommand*{\mcitethebibliography}{\thebibliography}
\csname @ifundefined\endcsname{endmcitethebibliography}
{\let\endmcitethebibliography\endthebibliography}{}


\begin{mcitethebibliography}{71}
\bibitem[pvs()]{pvs}
The author thanks Dan Frisbie for stimulating him to propose this alternative
  graphical representation (``PVS'') to the Fowler-Nordheim plots used in
  earlier works on transition voltage spectroscopy, because the latter may
  unphysically relate $V_t$ to a mechanistic transition \cite{Beebe:06} from a
  direct tunneling to a Fowler-Nordheim tunneling regime, which in fact does
  not occur at $V=V_t$. For this reason, the term ``Peak Voltage Spectroscopy''
  (PVS) suggested by the plots depicted in
  \figurename\ref{fig:generic-parabola}c, \ib{S2}c, \ib{S3}c, \ib{S4}c and
  \ref{fig:iv-NA}c, may be more appropriate than ``Transition Voltage
  Spectroscopy'' (TVS).\relax
\mciteBstWouldAddEndPunctfalse
\mciteSetBstMidEndSepPunct{\mcitedefaultmidpunct}
{}{\mcitedefaultseppunct}\relax
\EndOfBibitem
\end{mcitethebibliography}
\end{document}